\documentclass[journal=jacsat,manuscript=article]{achemso}

\usepackage[version=3]{mhchem} 
\usepackage{siunitx}
\usepackage{caption}
\usepackage{subcaption}
\usepackage{graphicx}
\usepackage{graphicx,color}
\usepackage{pgf,tikz}
\usepackage{chemfig}
\usepackage{amsmath}
\usepackage{multirow}
\usepackage[export]{adjustbox}
\usepackage{dcolumn}
\usepackage{bm}
\usepackage{multirow}
\usepackage{natbib}
\usepackage{etoolbox}
\usepackage{color}
\usepackage{xr}

\DeclareSIUnit\Molar{\textsc{m}}
\usepackage{atbegshi}
\AtBeginDocument{\AtBeginShipoutNext{\AtBeginShipoutDiscard}}
\author{Tobia Arcangeli}
\affiliation{Dipartimento di Scienze Molecolari e Nanosistemi, Università Ca' Foscari Venezia, Via Torino 155, 30172, Mestre, Venezia, Italy}
\alsoaffiliation{Dipartimento di Chimica, Materiali e Ingegneria Chimica "Giulio Natta", Politecnico di Milano, Sede Leonardo Edificio 6, Piazza Leonardo da Vinci 32, I-20133 Milano, Italy }
\author{Tatjana \v{S}krbi\'{c}}
\affiliation{Dipartimento di Scienze Molecolari e Nanosistemi, Università Ca' Foscari Venezia, Via Torino 155, 30172, Mestre, Venezia, Italy}
\author{Somiealo Azote}
\affiliation{Department of Physics and Bioinspired Institute, Syracuse University, Syracuse, NY 13244 USA}
\author{Davide Marcato}
\affiliation{Scuola Internazionale Superiore di Studi Avanzati (SISSA), Via Bonomea 265, 34136 Trieste, Italy}
\author{Angelo Rosa}
\affiliation{Scuola Internazionale Superiore di Studi Avanzati (SISSA), Via Bonomea 265, 34136 Trieste, Italy}
\author{Jayanth R. Banavar}
\affiliation{Department of Physics and Institute for Fundamental Science, 1258 University of Oregon, Eugene, OR 97403-1205 USA}
\author{Roberto Piazza}
\affiliation{Dipartimento di Chimica, Materiali e Ingegneria Chimica "Giulio Natta", Politecnico di Milano, Sede Leonardo Edificio 6, Piazza Leonardo da Vinci 32, I-20133 Milano, Italy }
\author{Amos Maritan}
\affiliation{Laboratory of Interdisciplinary Physics, Department of Physics and Astronomy ``G. Galilei'', University of Padova, Padova, Italy and INFN, Sezione di Padova, via Marzolo 8, 35131 Padova, Italy}
\author{Achille Giacometti}
\email{achille.giacometti@unive.it}
\affiliation{Dipartimento di Scienze Molecolari e Nanosistemi, Universit\`a Ca' Foscari Venezia, 30123 Venezia, Italy}
\alsoaffiliation{European Centre for Living Technology (ECLT) Ca' Bottacin, 3911 Dorsoduro Calle Crosera, 30123 Venezia, Italy}
 \title[]
  {Phase behaviour and self-assembly  \
   of semiflexible polymers  in poor-solvent solutions\footnote{Semiflexible polymers in solutions}}

\begin{document}

\begin{figure*}[htbp]
    \centering
    \captionsetup{justification=raggedright,width=1.0\linewidth}
\includegraphics[trim=0 0 0 0,clip,width=0.8\linewidth]{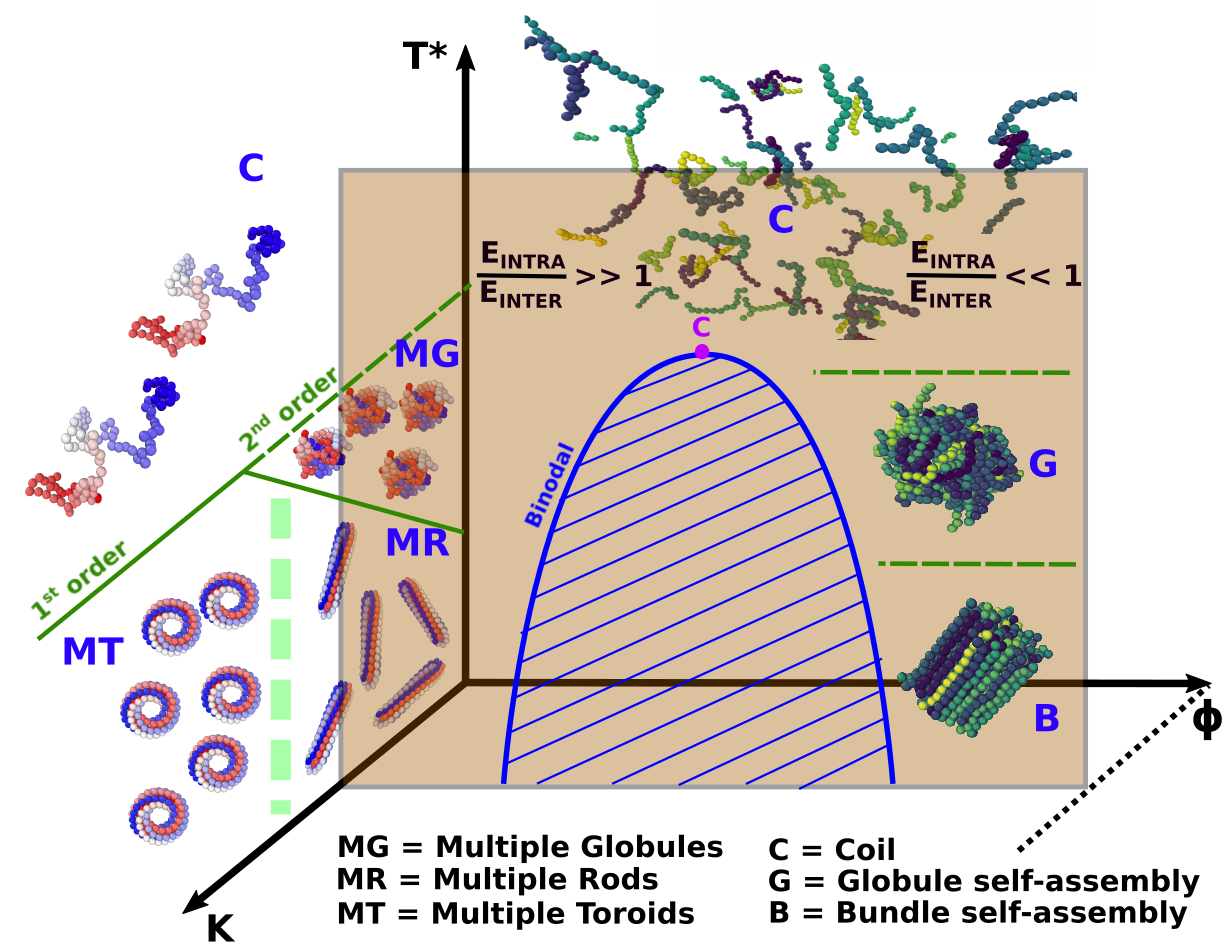}
    \caption*{Sketch of the proposed three dimensional phase diagram in the reduced temperature $T^{*}$-volume fraction $\phi$-stiffness $K$ space. In the  $T^{*}$-volume fraction $\phi$ plane for a fixed stiffness $K$, the binodal line ending at the critical point and separating the two different phases is shown. 
    }
    \label{fig:fig1}
\end{figure*}
\begin{abstract}
Using Langevin dynamics complemented by Wang-Landau Monte Carlo simulations, we study the phase behavior of single and multiple semiflexible polymer chains in solution under poor-solvent conditions. In the case of a single chain, we obtain the full phase diagram in the temperature-bending rigidity  (stiffness) plane and we provide connections with a classical mean field result on a lattice as well as with past results on the same model. At low bending rigidity and upon cooling, we find a second order coil-globule transition, followed by a subsequent first order globule-crystal transition at lower temperatures. The obtained crystals have the shape of a twisted rod whose length increases with the increase of the stiffness of the chain. Above a critical value of the stiffness, we also find a direct first order globule-crystal transition, with the crystal having the form of a twisted toroid. Close to the triple point, we find a region with isoenergetic structures with frequent switching from rods to toroids, with the toroid eventually becoming the only observed stable phase at higher stiffness. The model is then extended to many thermally equilibrated chains in a box  and the analogous phase diagram is deduced where the chains are observed to first fold into a globule bundle at low stiffness upon cooling, and then rearrange into a nematic bundle via a nucleation process involving an isotropic-nematic transition. As in the single chain counterpart, above a critical stiffness the chains are observed to undergo a direct transition from a gas of isotropically distributed chains to a nematic bundle as the temperature decreases in agreement with recent suggestions from mean field theory. The consequences of these findings for self-assembly of biopolymers in solutions are discussed.
\end{abstract}
\today
\clearpage

\maketitle

\section{Introduction}
\label{sec:introduction}
Many polymers, in particular biopolymers, are quite stiff \cite{Rubinstein2003}. Among these, semiflexible polymers are defined as polymers whose persistence length is significantly longer than the minimum length scale of the problem (e.g. the monomer size) yet still of the order of the total contour length \cite{Khokhlov2002}. 
Synthetic semiflexible polymers have interesting applications as scaffold that can be decorated with functional substituents, but semiflexible polymers find their most important applications in biopolymers such as collagen, DNA and the cytoskeletal components \cite{Gerrits2021}.
A typical example is a double strands DNA (dsDNA) that has a persistence length of 150 base pairs, corresponding to \SI{50}{\nano\metre}, a cross-section diameter of about \SI{2}{\nano\metre} and a contour length of $\approx$ \SI{1}{\micro\metre} \cite{Seol2007}. Notwithstanding its high electric charge, upon association with multivalent cation, it condenses into a compact form with characteristic morphologies that typically resemble either spheres or rods or toroids \cite{Vilfan2006}. In this case, the underlying physics can be  rationalized as a competition between solvent induced attraction between non-bonded monomers and an energetic penalty occurring upon bending.
The wormlike chain (WLC) model \cite{Rubinstein2003} (including volume interactions \cite{Montesi2004,Seaton2013,Marenz2016,Hoang2014,Skrbic2019,Majumder2021,Aierken2023jcp}) currently provides the best description of dsDNA elasticity for micron-sizes molecules \cite{Seol2007}, and hence serves as a suitable model for semiflexible polymers. 
In the WLC model, each nucleotide is represented by a spherical bead with a tethering potential mimicking the covalent bonds, a short range attraction complemented by excluded volume representing the effective monomer-monomer interactions induced by the solvent, and an elastic potential energy representing the energetic penalty associated with the bending of the chain thus providing stiffness~\cite{deGennes1972,Doi1988,Khokhlov2002,Rubinstein2003}. The latter also contributes to the persistence length of the polymer chain. \textcolor{black}{The WLC has also been successful in predicting spatial organization of DNA under confinement \cite{Curk2019}.}

Given the important consequences that the WLC model has on the understanding of thermodynamic and structural properties of semiflexible polymers, several studies  have attempted to define its phase diagram. As early as 1996, Doniach \textit{et al} \cite{Doniach1996} pioneered the qualitative key features of the phase diagram using a mean field theory on a lattice. This phase diagram was later confirmed and made more quantitative independently by Bastolla \& Grassberger \cite{Bastolla1997} and by Doye \textit{et al} \cite{Doye1998} using two different Monte Carlo simulations on a hypercubic lattice. These findings confirmed the mean field picture that the semiflexible chain upon cooling first folds into a structureless globule via a second order transition and then into a "crystalline" structure via a first order transition. Unlike the original mean field however, but in line with an improved Bethe approximation \cite{Lise1998}, the numerical simulations predicted a progressive thinning of the globule region upon increasing the chain stiffness until, eventually, a direct first order transition from a random coil to a crystal is observed. Due to the lattice structure constraint, here "crystal" means an ordered structure where the chain runs parallel for a number of lattice points that depend on the stiffness, makes a complete reversal, and runs again  aligned and antiparallel to the first direction. This process is then repeated along the neighbouring lattice layers over the full length of the chain. 

More recently, the phase behaviour of the WLC model has also been numerically studied off lattice. Montesi \textit{et al} \cite{Montesi2004} used Brownian dynamics and studied the low temperatures conformations; Seaton \textit{et al} \cite{Seaton2013} were the first to study the temperature-stiffness phase diagram using the Wang-Landau microcanonical method and to identify spheres, rods, and toroids as low density conformations. However, they also found additional structures whose stability was unclear. Marenz \textit{et al} \cite{Marenz2016} used essentially the same method and derived the same phase diagram with the addition of knotted structures, thus leaving the open question on what the actual stable ``ground state''  off lattice conformations are for this model. Simple energetic arguments \cite{Hoang2014,Hoang2015} supported by Monte Carlo simulations \cite{Hoang2014,Hoang2015,Aierken2023jcp,Aierken2023pre} seem to suggest that indeed only spheres, rods and toroids are stable structures, although the presence of knotted structures was confirmed by later studies \cite{Skrbic2019,Majumder2021}.
Additional numerical evidence on the stability of the rods and toroids \cite{Pham2005,Lappala2013,Kong2014,Dey2017,Wu2018role,Nguyen2023} also appears in line with experimental findings for DNA condensation \cite{Vilfan2006}.

In addition to single chain properties, the self-assembly of many semiflexible polymer chains in solution are of paramount importance in several contexts whose interest dates back to classic studies by Flory \cite{Flory1956a,Flory1956b}.
One of these is unquestionably peptide aggregation \cite{Aggeli2001}, where it was recently argued \cite{Fuxreiter2021} that under physiological conditions most proteins are capable of forming a condensed state with solid-like structure, known as the amyloid state, as well as another liquid-like condensed state denoted as the droplet state.
In this parallelism with conventional states of matter, the single protein native state provides the equivalent of a gaseous disordered state. Both the amyloid and the droplet states are formed via a nucleation process that drives a spontaneous self-assembly of the chains below a critical temperature.
Few numerical studies, with rather different models, have already identified signatures of this process. One study \cite{Nguyen2004} reports the observation of spontaneous fibril formation using molecular dynamics simulations of a coarse-grained model peptide. Here, the chain was flexible and the self-assembly was promoted by favourable interchain interactions. 
Another important set of studies \cite{Auer2007,Auer2008} used another model peptide known as ``thick polymer''~\cite{Hoang2004} and found similar results but without the use of additional interchain interactions.

The self-assembly of semiflexible polymer chains has also been studied before both experimentally \cite{Olmsted1998} and numerically \cite{Sheng1994,Sheng1996,Ivanov2003} but both experimental and computational limitations prevented the study of the full phase diagram. Recently, however, the original mean field on lattice for a single semiflexible chain \cite{Doniach1996} was extended to multichains \cite{Marcato2023} thus paving the way to an efficient numerical study. This work also provided a modern view to a combinatorial approach to the problem pioneered by Flory \cite{Flory1956a,Flory1956b}.

The aim of the present study is then twofold. First, we will revisit the single chain phase diagram in the temperature-stiffness plane and show that indeed it reproduces that previously found on the lattice with the only difference of having several low temperatures ``crystal phases'' with different morphologies (spheres, rods and toroids), rather than the single one allowed by the lattice structure.
Second, we extend this idea to multiple chains and study the same phase diagram at different concentrations by underscoring, at the same time, a nucleation process akin to that observed in amyloid formation.

The remaining of the paper is organized as follows.
Section~Theory and Methods provides the necessary theoretical and numerical tools to perform our analysis. Section Results for single chain reports results for the single semiflexible chain phase behaviour, whereas Section Results for multiple chains includes the multiple chains counterpart. Finally, Section Conclusions summarizes the findings of our study and puts them in perspective.

\section{Theory and Methods}
\label{sec:theory}
\subsection{Model potentials}
\label{subsec:model}
Consider a chain formed by $N$ spherical beads  of diameter $\sigma$ and mass $m$ located at positions $\mathbf{R}_{1},\ldots,\mathbf{R}_{N}$, and let $\mathbf{r}_{ij}=\mathbf{R}_j-\mathbf{R}_i$ be the vector distance between two beads. In this paper, we will study the phase behaviour of a single semiflexible polymer and the self-assembly properties of multiple such chains in solutions using Langevin dynamics (LD) supported by Wang-Landau Monte Carlo (MC) method.

For LD we consider a bead-spring polymer chain where consecutive monomers separated by a distance $r=\vert \mathbf{r}_{ij} \vert$ are connected by chemical bonds via the FENE potential \cite{Kremer1990}
\begin{eqnarray}
\label{sec1:eq1}
\phi_{\text{FENE}} \left(r \right) &=& -\frac{\kappa}{2} R_0^2 \ln \left[1- \left(\frac{r}{R_0} \right)^2\right] \qquad 0 \le r < R_0 \nonumber \\
&&
\end{eqnarray}
where $\kappa=30\epsilon/\sigma^2$ and $R_0=1.5\sigma$. 

\textcolor{black}{Two consecutive beads in the same chain interact via a Week-Chandler-Anderson Lennard-Jones shifted potential\cite{Weeks1971}:
\begin{eqnarray}
\label{sec1:eq1_1}
\phi_{\text{WCA}} \left(r \right) &=& 4 \epsilon \left[\left( \frac{\sigma}{r}\right)^{12} - \left( \frac{\sigma}{r}\right)^{6} + \frac{1}{4} \right] \qquad 0 < r \le 2^{1/6} \sigma
\end{eqnarray}
which is cut-off at $r=2^{1/6} \sigma$ as indicated.}
The choice of $\kappa = 30$ ensures that the average bond length is $0.97\sigma$ \cite{Auhl2003}.
The interactions between non bonded monomers separated by a distance $r$ are modeled with a truncated and shifted Lennard-Jones (LJ) potential \cite{Weeks1971}
\begin{eqnarray}
\label{sec1:eq2}
\phi_{\text{LJ}} \left(r \right) &=& 4 \epsilon \left[\left( \frac{\sigma}{r}\right)^{12} - \left( \frac{\sigma}{r}\right)^{6} - \left( \frac{\sigma}{r_c}\right)^{12} + \left( \frac{\sigma}{r_c}\right)^{6} \right] \nonumber \\
&&
\end{eqnarray}
for $0 < r \le r_c$, where $\epsilon$ defines the energy scale, $\sigma$ the length scale, and the cutoff is set to $r_c = 1.55 \sigma$ for reasons  explained below. 
All simulations were performed within a $NVT$ canonical ensemble using LAMMPS \cite{Plimpton1995,Thompson2022} complemented by in-house codes for pre- and post-processing. A Verlet algorithm \cite{Verlet1967} with a time step $\Delta t=0.005 \tau$ was used, in LJ units of time $\tau=\sqrt{m \sigma^2/\epsilon}$, with periodic boundary conditions applied in all directions. 
Temperature is controlled by a Langevin thermostat \cite{Schneider1978,Dunweg1991} with a dump coefficient equal to 1. All temperatures will be reported in reduced units $T^{*}=k_B T/\epsilon$, $k_B$ being the Boltzmann constant.

The bending rigidity is introduced via a standard bending worm-like chain (WLC) potential \cite{Rubinstein2003}
\begin{eqnarray}   
\label{sec1:eq3}
\phi_{\text{BEND}} \left(\theta \right)/k_B T &=& K\left(1+\cos\theta \right)
\end{eqnarray}
where $K$ is the (dimensionless) bending stiffness. We explore values of $K$ in the range $0 \le K \le 10$ with a resolution $\Delta K=0.2$ \textcolor{black}{in the case of a single chain and $\Delta K=1$ for the multi-chain case}.
The particular choice of the $\theta$ dependence in Eq.(\ref{sec1:eq3}) ensures elastic behavior at small $\theta$
as well as a flattening of this energy at larger angle. The total potential $\phi$ is then the sum of all terms given in Eqs.(\ref{sec1:eq1})-(\ref{sec1:eq3}).

The simulation protocol is as follows.
Preliminary consecutive annealing and equilibration runs is performed to generate equilibrated samples, followed by constant temperatures production runs used to accumulate statistics.    
In the annealing and equilibration the temperature of the Langevin thermostat was gradually lowered  in discrete steps of $\Delta T^* = 0.05$ for $T^*=2.00$ to $T^*=1.00$ and $\Delta T^* = 0.02$ in the range $T^*=1.00$ to $T^*=0.02$ corresponding to the region where condensed phases are observed.
For each temperature value reached in the cooling process, the system is allowed to equilibrate until no drift is observed in energy.

In the case of semiflexible polymers, the annealing procedure is carried out at fixed $K$ by decreasing the temperature of the system in finite steps, as previously outlined. 
The proper equilibration time is gauged to the specific final chain conformation and to the temperature condition. For high temperatures and non crystalline phases, an equilibration time $2 \times 10^7 \tau$ was found to be sufficient to equilibrate, whereas longer runs up to $3 \times 10^8 \tau$ were required in the globule-frozen transition region in order to observe the crystalline structure formation.
In all cases, the production run to collect data and perform statistical analysis was set to  $10^9 \tau$.

For MC simulations, we considered a tangent bead model where consecutive monomers are kept at distance $\sigma$ so that adjacent beads are tangent to one another \cite{Rubinstein2003}, and  non-consecutive beads separated by a distance $r$ are subject to a square-well potential
\begin{equation}
\label{sec1:eq4}
\phi_{SW}\left(r\right)=\begin{cases}
+\infty\,,\quad \,r < \sigma &\\
-\epsilon,\quad\, \sigma < r < \lambda \sigma&\\
0, \qquad r > \lambda \sigma&  .
\end{cases}
\end{equation}
 where $\lambda=1.2$ for reasons that will be clarified below.
 The semiflexible character of the chain was enforced via the same WLC potential $\phi_{BEND}(\theta)$ as in Eq.(\ref{sec1:eq3}).
 We use microcanonical Wang-Landau method \cite{Wang2001} to compute the zero-temperature `ground state' of the polymer, following the prescription given in a previous study by some of us \cite{Skrbic2019} as a benchmark for the MD simulations. In this scheme, the density of states  that are visited along the simulation was iteratively built, by filling consecutive energy histograms. The acceptance probability was chosen to promote moves towards less populated energy states, thus providing for increasing flatness of energy histograms with the length of the simulation and leading the search towards lower and lower energy states. Each time a configuration of lower total energy was recorded, it was saved in the trajectory for further analysis. The set of MC moves included both local-type moves, such as single-sphere crankshaft, reptation and end-point moves, as well as non-local-type moves, such as pivot, bond-bridging and back-bite moves. For every state point, between 2 and 5 ground-state trajectories were monitored, consisting of between 10$^9$ and 3$\times$10$^9$ MC steps per bead. For more details see Ref.\cite{Skrbic2019}. Results from MC simulations will be only used as cross-validation for the LD results. 

 In the case of multiple chain, LD $NVT$ simulations are extended to $N_c$ identical chains and their self-assembly properties are studied as a function of the temperature $T$ and the bending rigidity $K$. In this case, we use $N_c=80$ a short polymer chain of $N=12$ monomers, and scale the volume to achieve a specific molar concentration. In this case, equilibration runs lasted at least $4 \times 10^{7} \tau$ and production runs lasted up to  $4 \times 10^{8} \tau$. 

\subsection{Order parameters}
\label{subsec:order}
In order to distinguish different phases and chain morphologies, we use both thermodynamic and structural order parameters. In doing this we will try to make clear what are thermodynamic transition, related to discontinuities in some thermodynamic parameter, and what are structural transitions, related to a morphological change not associated with any discontinuity in the thermodynamics.

In $NVT$ ensemble thermodynamics transition are identified as peaks in the specific heat
\begin{eqnarray}
\label{sec1:eq5}
    C(T)=\frac{ \left \langle E^2 \right \rangle -\left \langle E \right \rangle ^2}{k_{B}T^2}
\end{eqnarray}
where  $\langle E^2 \rangle $ and $\langle E \rangle$ are thermal averages measured in MD simulations as temporal averages over the total energy trajectory provided by production runs on equilibrated samples.
The mean-square radius of gyration 
\begin{eqnarray}
\label{sec1:eq6}
    \left \langle R_g^2\left(T\right) \right \rangle &=&\frac{1}{N}\sum_{i=1}^N\left(\mathbf{R}_i - \mathbf{R}_{CM} \right)^2 
\end{eqnarray}
is used as order parameter to distinguish the coil phase from the condensed phases. Here $\mathbf{R}_{CM}$ is the center-of-mass of the chain, and in the case of multiple chains, an obvious generalization is considered. 

Different morphologies will be identified by providing structural order parameters based on the distance $R_{ij}=\vert \mathbf{R}_{i}-\mathbf{R}_{j} \vert$ that clearly distinguish between a rod-like and a toroidal-like structure, as will be further elaborated below. In addition, the moment of inertia tensor of a single chain
\begin{eqnarray}
\label{sec1:eq7}
\mathbf{I}_{\mu \nu} &=& \sum_{i=1}^{N} m_{i} \left(R_{i}^2 \delta^{\mu \nu}- R_{i}^{\mu} R_{i}^{\nu} \right)
\end{eqnarray}
which, as we shall see, is a well defined measure of the shape of the collapsed chain. Here, $\mu,\nu=1,2,3$ are the cartesian indices. In addition, a globular crystal structure can be easily discriminated from an unstructured globule by characteristic peaks in the radial distribution function $g(r)$ \cite{Hansen2001}. 

In the case of multiple chains, an important signature of the self-assembly is the difference between intra-chain potential energy $E_{intra}$ and its inter-chain counterpart $E_{inter}$. On defining  $\phi_{c, c^{\prime}} (ij)$ as the full pair interactions $\phi_{\text{LJ}}(ij)+\phi_{\text{BEND}}(ij)$ (Eqs.(\ref{sec1:eq2}) and (\ref{sec1:eq3})) between monomers $i$ and $j$ in the $c$-th and $c^{\prime}$-th chains respectively, we have 
\begin{eqnarray}
\label{sec1:eq8a}
 E_{intra} &=& \frac{1}{2} \sum_{c=1}^{N_c} \sum_{i=1}^{N} \sum_{j \ne i=1}^{N}\phi_{c c} \left(ij\right) \\
 \label{sec1:eq8b}
 E_{inter} &=& \frac{1}{2}\sum_{c=1}^{N_c} \sum_{c^{\prime}\ne c=1}^{N_c} \sum_{i=1}^{N} \sum_{j=1}^{N} \phi_{c c^{\prime}} \left(ij\right)
\end{eqnarray}
The total potential energy $E$ will then be the sum of the above two terms, but the relative balance of $E_{intra}$ and $E_{inter}$ will be very different in the the case of independent collapsing, in which case $E \approx E_{intra}$ and $E_{inter} \approx 0$, and in the case of self-assembly, in which case $E_{inter} \gg E_{intra}$. In $E_{intra}$ defined in Eq.(\ref{sec1:eq8a}) the factor $j=i$ was clearly omitted  and the factor $1/2$ was included to avoid double counting. Likewise, in $E_{inter}$ the factor $c^{\prime}=c$ is omitted and the factor $1/2$ was included for the same reasons.
When $K>0$, we expect the onset of a nematic phase with different chains aligning parallel to one another in a close bundle, and this can be achieved by evaluating the Veilliard-Baron tensor \cite{Vieillard1974} 
\begin{eqnarray}
  \label{sec1:eq9}
  \mathbf{Q}_{\widehat{\mathbf{u}}} &=& \left \langle \frac{1}{N_c} \sum_{c=1}^{N_c} \left[\frac{3}{2} \widehat{\mathbf{u}}_{c} \widehat{\mathbf{u}}_{c} -\frac{1}{2} \mathbf{I} \right]  \right \rangle
\end{eqnarray}
where $\widehat{\mathbf{u}}_c$ is the unit vector identifying the orientation of the $c-$th chain (assumed rod-like shaped) in space.
 The maximum eigenvalue $S$ gives the nematic order parameter $\approx 1$ for a nematic phase and $\approx 0$ for an isotropic disordered phase, and the corresponding eigenvector gives the nematic director $\widehat{\mathbf{N}}$ of orientation of the bundle. In principle, this order parameter could be also exploited in the case of a single albeit sufficiently long chain.

\section{Results for single chain}
\label{sec:single}
\subsection{Matching the square-well and the LJ potentials via the second virial coefficients}
\label{subsec:matching}
 In this part, we justify the choice $r_c=1.55\sigma$ as cut-off value for the LJ potential given in Eq.(\ref{sec1:eq2}). Our guideline was to be able to reproduce as closely as possible the phase diagram for the flexible polymer ($K=0$) as obtained by Taylor et al. \cite{Taylor2009} via the Wang-Landau \cite{Wang2001} method. This matching will be further supported by our own calculation using this method that will be discussed in the next subsection.
The SW potential Eq.(\ref{sec1:eq4}) depends on $\sigma$, $\lambda$, and $\epsilon$ parameters. The LJ potential Eq.(\ref{sec1:eq2}) depends on $\sigma$, $\epsilon$ and $r_c$ parameters. Assuming identical lengths ($\sigma$) and energy ($\epsilon$) scales, we essentially need a relation between $\lambda$ and $r_c$ at a given temperature. This can be achieved by matching the second virial coefficients 

\begin{eqnarray}
\label{sec2:eq1}
  B_{2}^{SW}\left(\sigma,\epsilon,\lambda,T\right) &=& B_{2}^{LJ}\left(\sigma,\epsilon,r_c,T\right)
\end{eqnarray}
where the SW case can be readily evaluated as
\begin{eqnarray}
\label{sec2:eq2}
  B_{2}^{SW}(\sigma,\epsilon,\lambda,T) &=&\frac{2\pi}{3}\sigma^3\left[\left(1-\lambda^3\right)e^{-\beta\epsilon}-\lambda^3\right] \nonumber \\
  &&
\end{eqnarray}
whereas the LJ second virial coefficient
\begin{eqnarray}
\label{sec2:eq3}
  B_{2}^{LJ}(\sigma,\epsilon,r_c,T)&=&-\frac{1}{2}\int_{0}^{r_c} dr~4\pi r^2 \left[e^{-\beta\phi_{LJ}(r)}-1\right] \nonumber \\
  &&
\end{eqnarray}
is calculated numerically although an analytical solution does exist \cite{Gonzalez2015}.
\begin{figure}[htbp]
    \centering
    \captionsetup{justification=raggedright,width=1.0\linewidth}
\includegraphics[trim=0 0 0 0,clip,width=0.8\linewidth]{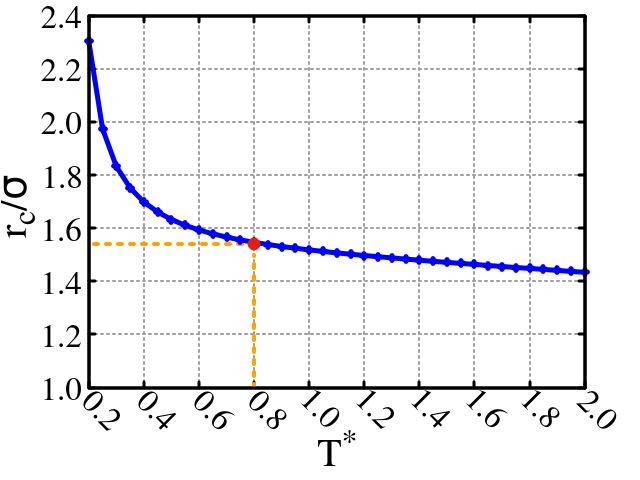}
    \caption{Effective $r_c(T^*)$ corresponding to $\lambda = 1.18$ computed for each temperature point in the range $T^{*}=0.2-2.0$. Highlighted is a point corresponding to $T^{*}=0.8$ and $r_c/\sigma=1.55$ that has been selected as the optimal value (see text)}
    \label{fig:fig1}
\end{figure}
For a flexible ($K=0$) chain having the same number of monomers $N=128$ as in Ref. \cite{Taylor2009}, Eq.(\ref{sec2:eq1}) was solved numerically to find the optimal LJ cut-off $r_c$ that matches the SW width value $\lambda=1.18$ in Ref. \cite{Taylor2009}and clearly leads to a double transition from expanded coil to collapsed globule to frozen crystallite (see Figure 6 in Ref. \cite{Taylor2009}) with a temperature range $0.2 \le T^{*} \le 2.0$. 
For a $\lambda = 1.18$ system, the WL analysis predicts a frozen crystal-globule phase transition at $T_{f,g}=0.493(4)$ and globule-coil transition at the critical temperature $T_{g,c}=0.800(2)$ (see Table II in Ref. \cite{Taylor2009}). Figure \ref{fig:fig1} reports the results of this calculation in which the point $T^{*}=0.8$ and $r_c/\sigma=1.55$ has been highlighted because it is found to reproduce reasonably well the aforementioned thermodynamic behaviour, with the two transitions occurring nearly at the same temperatures observed in Ref. \cite{Taylor2009}.

\subsection{Comparison with Wang-Landau results}
\label{subsec:comparison}
 
\begin{figure*}[htpb]
\centering
\begin{subfigure}[b]{0.30\textwidth}
    \centering
    \includegraphics[scale=0.35]{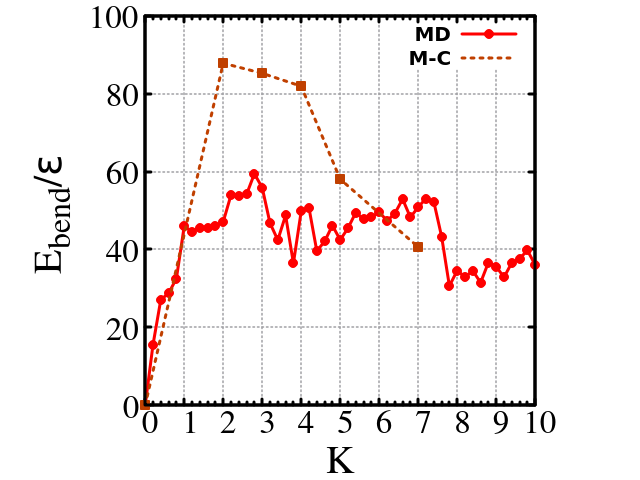}
     \caption{}\label{fig:fig2a}
    \end{subfigure}
    \begin{subfigure}[b]{0.30\textwidth}
    \centering
    \includegraphics[scale=0.35]{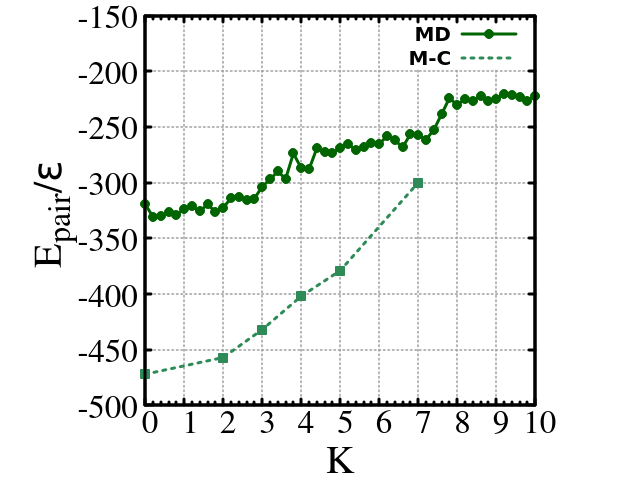}
     \caption{}\label{fig:fig2b}
    \end{subfigure}
    \begin{subfigure}[b]{0.30\textwidth}
    \centering
    \includegraphics[scale=0.35]{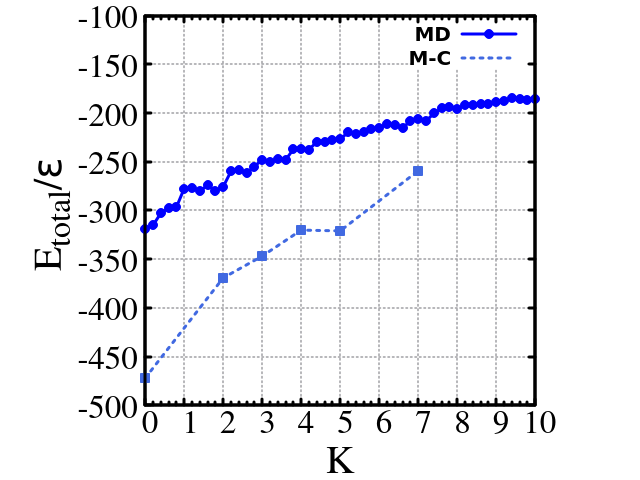}
     \caption{}\label{fig:fig2c}
    \end{subfigure}
\caption{Comparison between ground state energies as obtained by Wang-Landau method \cite{Skrbic2019} and present LD at the lowest considered temperature $T^{*}=0.02$. (a) Comparison of the total (dimensionless) bending energy $E_{bend}/\epsilon$ as a function of the bending rigidity $K$; (b) Comparison of the total (dimensionless) pair energy $E_{bend}/\epsilon$ as a function of the bending rigidity $K$; (c) Comparison of the total (dimensionless) energy $E_{total}/\epsilon$ as a function of the bending rigidity $K$. }
\label{fig:fig2}
\end{figure*}

To provide additional support to the correctness of our LD simulations, we here perform a comparative study with the different ground state energies obtained using the Wang-Landau MC method \cite{Skrbic2019}. In the LD case, we consider the energies corresponding to the lowest considered temperature $T^{*}=0.02$, whereas for WL we used $\lambda=1.20$ which is  nearly identical to the value used in previous Section \ref{subsec:matching} to parametrize the LD calculation.

Figure \ref{fig:fig2} reports the results of this comparison, divided as a comparison of the total bending energy $E_{bending}/\epsilon$ (Figure \ref{fig:fig2a}), total pair energy $E_{pair}/\epsilon$ (Figure \ref{fig:fig2b}), and the total sum of the two $E_{total}/\epsilon$ (Figure \ref{fig:fig2c}). Note that for LD $E_{bending}/\epsilon$ considers the sum of all terms with pair interactions given by Eq.(\ref{sec1:eq3}) and $E_{pair}/\epsilon$ the sum of all terms with pair interactions given by Eq.(\ref{sec1:eq2}). By contrast, in WL they are the sum of all pair interactions given by Eq.(\ref{sec1:eq3}) as before and the sum of all pair interactions given by Eq.(\ref{sec1:eq4}) respectively. In both cases, these are the non-bonded interaction energies, but the specific meaning in each case is different. Moreover, in principle the WL method provides a very precise measure of the ground state energy, whereas in the LD case, it depends on the selected temperature and this choice is rather arbitrary. In view of these differences, the values and the trend stemming from the two analyses are not inconsistent. While for pair interactions the WL energies provide a lower bound as expected (Figure \ref{fig:fig2b}), for bending interactions the WL energies are found to follow the same initial path, overcome the LD counterpart, and then return back to approximately the same LD values (Figure \ref{fig:fig2a}). These discrepancies are likely to be ascribed to the different implementation of the bending in the model used for WL calculations \cite{Skrbic2016a,Skrbic2019}.

We further note that the total pair energy $E_{pair}/\epsilon$  is always approximately an order of magnitude higher (in magnitude) as compared with the total bending energy $E_{bending}/\epsilon$, and that the latter appears to flatten out for $K \approx 10$.

Our LD findings may be compared with those reported in Ref. \cite{Aierken2023jcp} who also performed a similar analysis and derived nearly identical results (see their Figure 1).

\subsection{Specific heat and Radius of gyration}
\label{subsec:specific}
We performed extensive LD simulations at different temperatures $T^{*}$  and bending rigidity $K$ to identify the phase behavior of the single semiflexible chain.
\begin{figure*}[htbp]
   \begin{subfigure}[b]{8cm}
    \includegraphics[trim= 0 0 0 0,clip,width=1.0\linewidth]{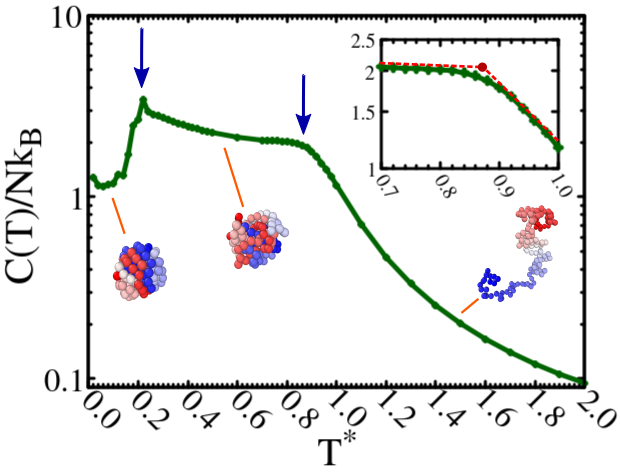}
    \caption{}\label{fig:fig3a}
    \end{subfigure}
    \hfill
    \begin{subfigure}[b]{8cm}
    \includegraphics[trim= 0 0 0 0,clip,width=1.0\linewidth]{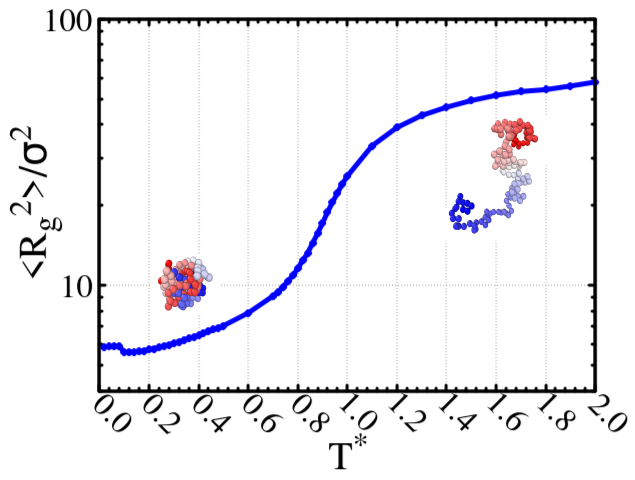}
    \caption{ }\label{fig:fig3b}
   \end{subfigure}
  \caption{(a) Specific Heat per monomer as a function of the reduced temperature $T^{*}=k_B T/\epsilon$ for the flexible chain case $K = 0$ and $r_c/\sigma = 1.55$. The peak at $T_{fg}^{*}=0.22$, indicated by the left arrow corresponds to the globule-crystal transition. The kink indicated by the second arrow and highlighted in the inset, identifies the coil-globule transition at $T_{gc}^{*}=0.87$; (b) Dimensionless mean square radius of gyration  $\langle R_g^2 \rangle$ as a function of the reduced temperature $T^*$. In both (a) and (b) the typical high temperature swollen and the low temperature globule/crystal conformations are shown.}
  \label{fig:fig3}
\end{figure*}
As anticipated, the phase behavior will be assessed using a combination of order parameters that will allow us to locate the phase transitions. In particular, a combination of the results for the specific heat $C(T)$ and of the mean squared gyration radius $\langle R_g^2 \rangle$ can identify both the coil-to-globule and globule-to-crystal phases in the case of the fully flexible ($K=0$) chain \cite{Taylor2009}. These are the analogs of the gas-liquid and liquid-solid transitions in simple liquids \cite{Paul2007}.
This is reported in Figure \ref{fig:fig3} where the behavior of the specific heat per monomer $C(T)/N k_B$ (Figure \ref{fig:fig3a}) and of the mean square radius of gyration $\langle R_g^2 \rangle$ (Figure \ref{fig:fig3b}) are computed as a function of the reduced temperature $T^{*} = k_B T/\epsilon$ using the cut-off value $r_c/\sigma=1.55$ obtained in Section on the matching of the second virial coefficient \ref{subsec:matching}. The peak in the specific heat occurring at $T_{fg}^{*}=0.22$ (left arrow) marks the globule-crystal transition, whereas the kink occurring at  $T_{gc}^{*}=0.87$ (right arrow) identifies the coil-globule transition. In this latter case, the transition temperature has been estimated in the point of maximum curvature identified by the intersection of the two regression lines shown in red in the inset. The shape of $C(T)$ is consistent with the behaviour obtained via the WL method by Taylor et al \cite{Taylor2009} with a SW with $\lambda=1.18$ (see Table II in Ref. \cite{Taylor2009} and discussion in Section on the matching of the second virial coefficient \ref{subsec:matching}) which predicts a globule-coil transition temperature of $T_{gc}^{*} = 0.800$. A slightly worse agreement is observed for the globule-crystal transition which is found at $T_{fg}^{*} = 0.493$. The collapse of the chain is also indicated by the behavior of the mean square radius of gyration $\langle R_g^2\rangle$ which is observed to have a marked decrease below a reduced temperature $T^{*}=k_B T/\epsilon \approx 1$. 

The above phase behaviour for the flexible $K=0$ case is also confirmed by the radial distribution function $g(r)$ that provides evidence of the gradual onset of crystalline ordering upon cooling as Bragg peaks as illustrated in Supplementary Section S and Supplementary Figure SIII. 

\begin{figure*}[htbp]
   \begin{subfigure}[b]{8cm}
    \includegraphics[trim= 0 0 0 0,clip,width=1\linewidth]{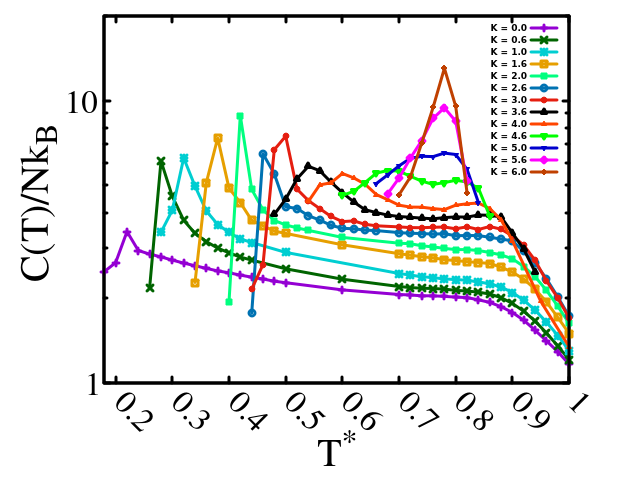}
    \caption{}\label{fig:fig4a}
    \end{subfigure}
    \hfill
    \begin{subfigure}[b]{8cm}
    \includegraphics[trim= 0 0 0 0,clip,width=1.0\linewidth]{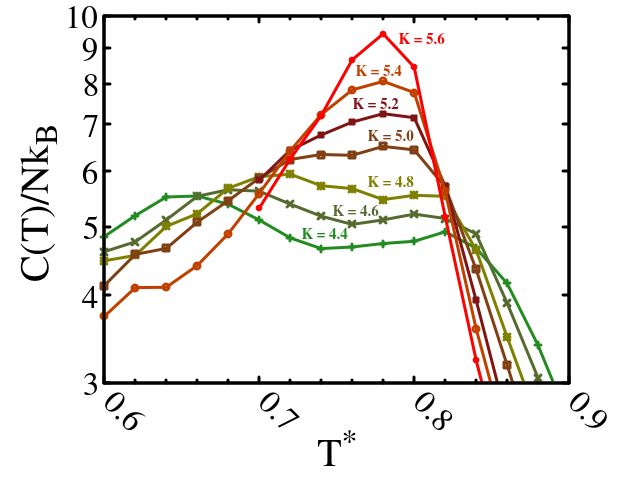}
    \caption{ }\label{fig:fig4b}
   \end{subfigure}
  \caption{(a) Specific heat per monomer $C(T)/Nk_B$ vs temperature $T^*$ for $K=1$ to $K=6$. The behaviour obtained for  $K=1,2,3,4$ (green lines) shows the presence of two critical temperatures corresponding to the crystal-globule transition (first highest peak) and globule-coil transition (kink at high temperature), as already seen for the flexible chain case. The globule phase disappears when a critical stiffness of $K=5$ is reached (brown line); for bigger values of $K$ the regime is characterized by single first order transition between the coil phase and an the ordered phase, as shown for the case $K=6$ (red line). (b) Enlarged view of the collapsing of the two critical peaks around the triple point with stiffness values from $K=4.4$ to $K=5.6$.}
  \label{fig:fig4}
\end{figure*}

The same analysis can be extended to the case of a semiflexible polymer ($K>0$). This is reported in Figure \ref{fig:fig4} and it underscores a novel feature that emerges in this regime. The $K=0$ has been included as a benchmark.
As $K$ increases, the marked peak occurring at low temperature signaling the globule-crystal transition gradually shifts to higher temperatures, whereas the kink occurring at higher temperature associated with the coil-globule becomes more and more evident without any significant shift in its location. 
Hence, in this range, upon cooling we still observe first a second order coil-globule transition, followed by a first order globule-crystal transition at lower temperature.
This trend persists until a critical value $K_c \approx 5$ above which the globule-crystal transition peak merges with the coil-globule transition peak. These findings can be interpreted as follows. In the region $0 \le K \le K_c \approx 5$, the collapse of the polymer occurs in two steps: a second order transition from a coil to a globule, followed by a first order transition from a globule to a crystal. In contrast, above $K_c \approx 5$, a single, direct, first order transition occurs from a coil to a crystal phase. 
Supplementary Figure SIV provides evidence of the onset of a bimodal distribution at higher bending rigidity ($K=7$-$10$) which occurs at nearly constant temperature $T^{*} \approx 0.80$, although a dedicated calculation would be required to fully elucidate this particular feature. 

Doniach \textit{et al} \cite{Doniach1996} first provided a detailed phase diagram in the temperature-bending rigidity plane of a semiflexible polymer chain on a lattice using a mean field approach.
They clearly identified the two coil-globule and globule-crystal transition and their different second/first order, but they were unable to find the merging of the two transitions into a single one for sufficiently large stiffness.
This deficiency can be overcome using a Bethe approximation \cite{Lise1998}, and numerical support to this phase diagram on the lattice came from both \citeauthor{Bastolla1997} \cite{Bastolla1997} and by Doye \textit{et al} \cite{Doye1998}.
Numerical work off-lattice does exist as well. Seaton \textit{et al} \cite{Seaton2013} used a two-dimensional variant of the Wang-Landau microcanonical formalism to provide a detailed description of the transition from a flexible to stiff polymer chain; another study \cite{Marenz2016,Majumder2021} focused more on the onset of knotted phases; \v{S}krbi\'{c} \textit{et al} \cite{Skrbic2016a,Skrbic2019} also studied the phase diagram of semiflexible polymer chains  along with a comparison with another type of stiffness, denoted as ``entropic stiffness'', originating from excluded volume.
Finally, recent work from two different groups \cite{Du2022,Aierken2023pre,Aierken2023jcp,Aierken2023pccp} also provided a detailed discussion on the phase behaviour of a very similar model.
Although a phase diagram is present in all the above studies, in one way or another, the details are different, and the fact that the models differ in some minor details does not help the physical interpretation. It is not always clear how to discriminate between stable intermediate phases, and metastable phases originating from kinetic trapping. For this reason, in the next Section we will provide our own phase diagram where we will be guided by both detailed numerical findings and by physical insights given by the mean field theories alluded earlier as well as by experiments.
\subsection{Phase diagram for a single chain}
\label{subsec:phase_single}
\begin{figure*}[htbp]
\centering
\includegraphics[width=16.0cm]{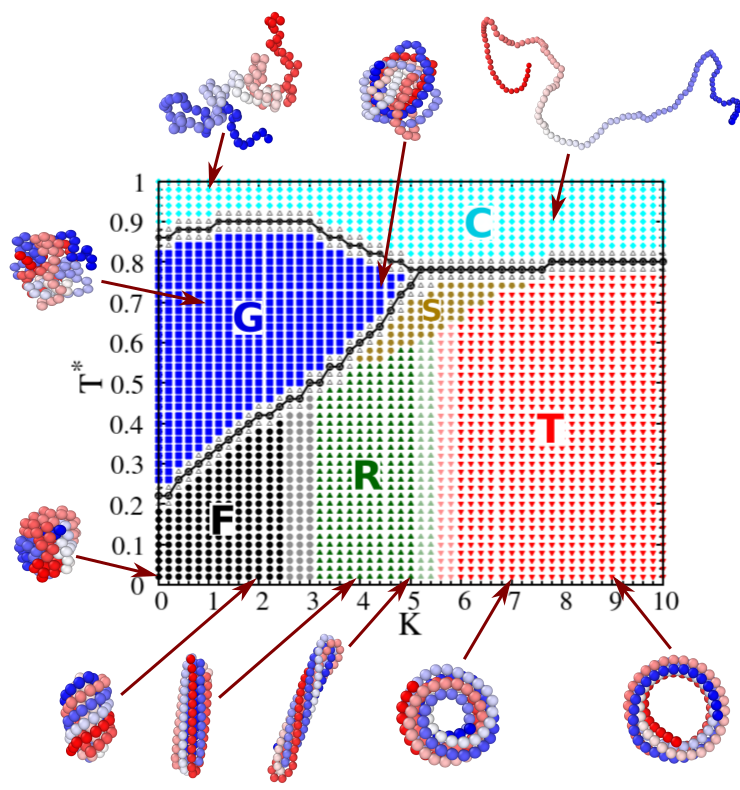}
\caption{Phase diagram of the single chain configurations in the reduced temperature $T^{*}$-bending rigidity $K$ plane. Different phases are identified with different colours and different shapes.
Toroid (\textbf{T} red triangles down), Rod (\textbf{R} green triangles up), Frozen-crystal/yarn (\textbf{F} black circles), Globule/disordered condensed phase (\textbf{G} blue squares), Coil (cyan diamond \textbf{C}), Switching degenerate region (\textbf{S} brown circles).
Black solid line identify the coil-globule, globule-crystal and coil-crystal regions, according to the behaviour of the specific heat $C(T)$.
In the faded regions $2.6 \leq K \leq 3.0$ and $5.2 \leq K \leq 5.8$ more than one conformation has been obtained during multiple annealing experiments at the same $K$.}
\label{fig:fig5}
\end{figure*}

Figure \ref{fig:fig5} presents our phase diagram in the temperature-bending rigidity plane, drawn according to the scenario we have been discussing in the previous Subsection.
Consider the region $0\le K \le K_c \approx 5$.
On cooling from high temperature, one first observe a coil-globule transition identified by a kink in the specific heat, and then a globule-crystal transition identified by a second, more marked peak in the specific heat (see Figure \ref{fig:fig4a}).
By contrast, for $K > K_c \approx 5$, a direct coil-crystal transition is observed (Figure \ref{fig:fig4b}), in very close analogy to what is found numerically on the lattice \cite{Bastolla1997,Doye1998} and predicted by mean field theory \cite{Doniach1996,Lise1998}.
Unlike results of on-lattice simulations, however, two different conformations are observed at low temperatures \cite{Hoang2014,Hoang2015}.
At low $K\le 2.2$, the chain starts from the globular spherical shape characteristic of the flexible ($K=0$) counterpart, and gradually tends to form a rod-like structure whose length is approximately equal to the persistence length $L_p \propto K/T^{*}$ and increases with $K$ and decreases with $T^{*}$ \cite{Auhl2003}.
It also tends to wrap and twist around this first rod-like structure to maximize the number of favourable contacts. This elongation incurs an energetic penalty (see Figure \ref{fig:fig2a}) with no increase in favorable contacts (see Figure \ref{fig:fig2b}), thus leading to an increase of the total energy (Figure \ref{fig:fig2c}).
A glance at the phase diagram shows that this is the region labelled as \textbf{F} where the chain has a yarn-like shape.
A clear rod-like structure clearly emerges for $2.2 < K \le K_c \approx 5$ and this is the region labelled as \textbf{R} in the phase diagram of Figure \ref{fig:fig5}.
Here, however, the increase in the total energy can be ascribed to the decrease of the number of favourable contacts rather than to the increase of the bending stiffness, as this latter flattens out (Figure \ref{fig:fig2a}).
This feature was also observed in Ref. \cite{Aierken2023jcp}. Here, we note that we do not regard \textbf{F} as a different phase from \textbf{R} although it is highlighted in Figure \ref{fig:fig5} just because here the structure is sphere-like with an aspect ratio not significantly larger than 1.
We will elaborate on this point further below. By contrast, we consider the flexible limit case $K=0$ as "singular" because its ground state is highly degenerate.

For $K> K_c \approx 5$, the bending energy fluctuates around a constant value (Figure \ref{fig:fig2a}) whereas the number of favourable contacts gradually starts to decrease (Figure \ref{fig:fig2b}) with no apparent structural discontinuity. As further elaborated below, however, a toroidal-like conformation is always observed in this region, whose inner radius increases with $K$ (see snapshots in Figure \ref{fig:fig5}). This structural change is abrupt and it can be rationalized with simple energetic arguments \cite{Hoang2014,Hoang2015}. As the radius of the toroid increases, the number of favourable contacts decreases but the energetic penalty due to bending also decreases (Figure \ref{fig:fig2a}), thus making the total energy increase gradually (Figure \ref{fig:fig2c}). Again, a similar argument appears in Ref. \cite{Aierken2023jcp}. 

A note of caution is in order here. As alluded earlier, the phase diagram of Figure \ref{fig:fig5} was considered before by few studies  \cite{Seaton2013,Marenz2016,Majumder2021,Du2022,Aierken2023pre} but there is no general consensus on the details of the various phases and their stability in the thermodynamic limit.
For instance, in Ref. \cite{Seaton2013} an additional ``hairpin'' conformation, a combination of toroid and rod, was predicted as a stable low temperature structure.
A similar prediction was also observed by the authors of Refs. \cite{Marenz2016} and \cite{Du2022} who also predicted the existence of additional structures such as knotted ones \cite{Marenz2016} or double helices \cite{Du2022}.
However, both these predictions are based on numerical simulations of much shorter polymer chains ($\le 30$ monomers in all cases) compared to those of the present study (128 monomers). Much longer polymer chains ($\ge 1000$ monomers) were considered in Ref. \cite{Lappala2013} for some state points, but the full phase diagram was not studied. Additional more recent results based on simulations with 55 monomers do not provide evidence of stable hairpin low temperatures structures \cite{Aierken2023pre,Aierken2023jcp}. We also observed the onset of a hairpin conformation in some of our runs, but this conformation was not found to be stable under a change in the temperature, and this is the reason why it is not included in the phase diagram of Figure \ref{fig:fig5}. An additional reason hinges on a physical basis. Although it is true that there is a transition region between rods and toroids (the vertical faded region between \textbf{R} and \textbf{T} region) where the two conformations are nearly isoenergetic (see also Ref. \cite{Aierken2023jcp}), the minimization of the energy in the two cases appears to be rather different \cite{Hoang2014,Hoang2015} and hence it is hard to contemplate to a stable structure formed by a combination of the two. 

In summary, our present findings support the existence of  rods and toroids as stable low temperatures conformations (\textbf{R} and toroid \textbf{T} in Figure \ref{fig:fig5}) as a natural extension of the single rod-like ground state found from numerical simulations on lattice \cite{Bastolla1997,Doye1998}, and this is supported by previous energetic arguments \cite{Hoang2014,Hoang2015}. It is also in accord to a significant extent with past similar  results \cite{Aierken2023jcp}. None of the above studies, however, underscored the existence of two different regimes associated with a double transition (at low $K$) and single transition (at high $K$) \cite{Aierken2023pre}. \textcolor{black}{Moreover, the phase diagram of Figure \ref{fig:fig5} correctly predicts that the critical temperature of the  first-order globule-crystal transition is positively sloped as a function of the stiffness. This results from the decreasing entropy (due to the stiffening) combined with the increasing energy (due the the decreasing number of favorable contacts), in complete agreement with mean-field theories \cite{Doniach1996,Lise1998} and numerical simulations on lattice \cite{Bastolla1997,Doye1998}.}
\begin{figure*}[htbp]
\centering
\begin{subfigure}[b]{0.40\textwidth}
    \centering
    \includegraphics[width=7.0cm]{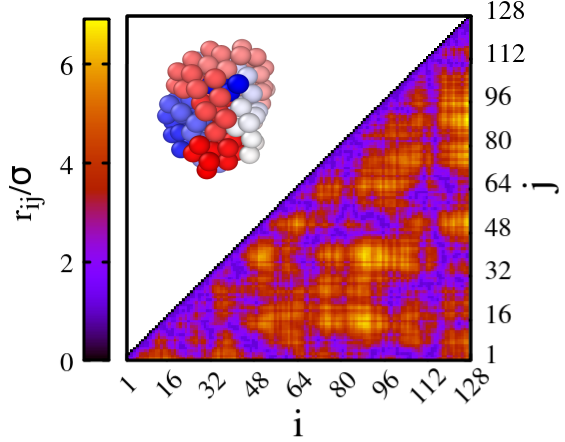}
     \caption{}\label{fig:fig6a}
    \end{subfigure}
    \qquad \qquad \qquad
    \begin{subfigure}[b]{0.40\textwidth}
    \centering
    \includegraphics[width=7.0cm]{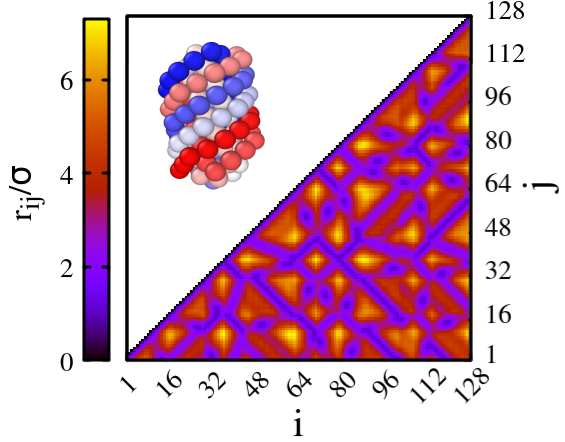}
     \caption{}\label{fig:fig6b}
    \end{subfigure}
    \\
    \begin{subfigure}[b]{0.40\textwidth}
    \centering
    \includegraphics[width=7.0cm]{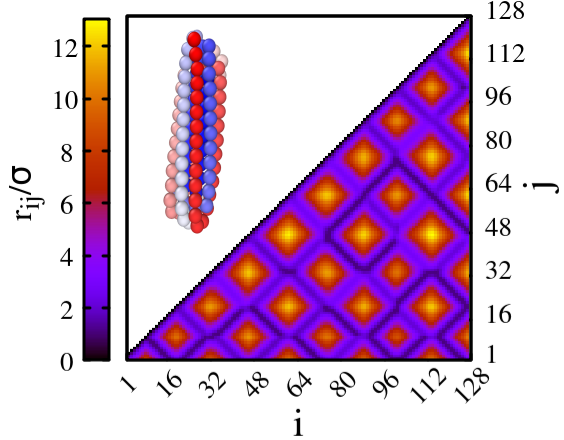}
     \caption{}\label{fig:fig6c}
    \end{subfigure}
    \qquad \qquad \qquad
    \begin{subfigure}[b]{0.40\textwidth}
    \centering
    \includegraphics[width=7.0cm]{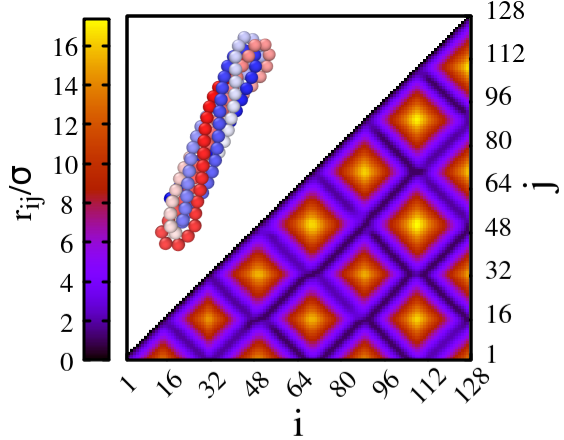}
     \caption{}\label{fig:fig6d}
    \end{subfigure}
    \\
    \begin{subfigure}[b]{0.40\textwidth}
    \centering
    \includegraphics[width=7.0cm]{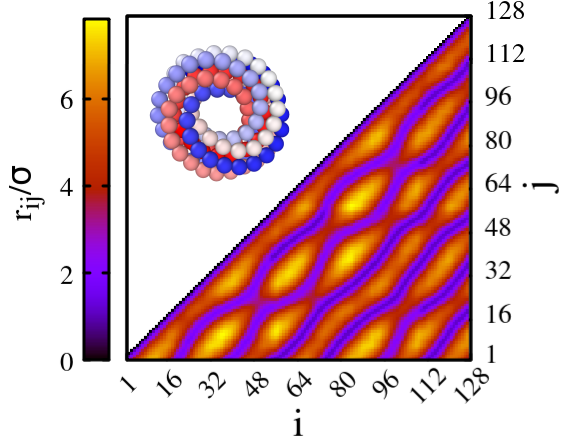}
     \caption{}\label{fig:fig6e}
    \end{subfigure}
    \qquad \qquad \qquad
    \begin{subfigure}[b]{0.40\textwidth}
    \centering
    \includegraphics[width=7.0cm]{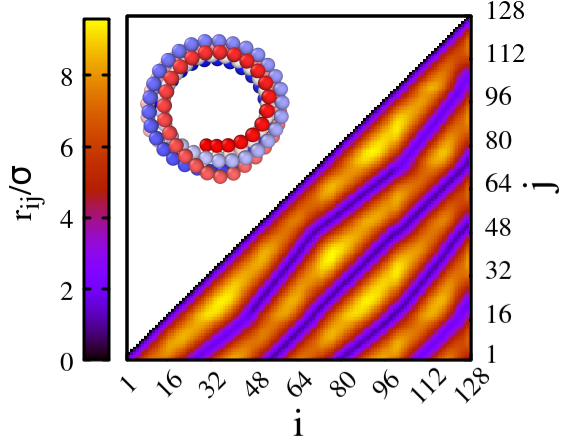}
     \caption{}\label{fig:fig6f}    
    \end{subfigure}
  \caption{Contact map of selected crystalline configurations at the lowest considered temperature $T^* = 0.02$ and for increasing bending rigidity $K$. (a) Globule-like crystal ($K = 0$); (b) Twisted-yarn-like crystal ($K = 2$); (c) Short-twisted rod ($K = 4$); (d) Long-twisted rod ($K = 5$); (e) Small toroid ($K = 6$); (f) Large toroid ($K = 9$). The contour plot reports all monomers from one end to the other only for $i,j=1,\ldots,128$, $i<j$ in view of the symmetry under the $i \longleftrightarrow j$ interchange.} 
  \label{fig:fig6}
\end{figure*}
As anticipated in Section dedicated to order parameters \ref{subsec:order}, we can distinguish between rods and toroids also on the basis of the contact map analysis, similarly to some previous studies \cite{Montesi2004,Aierken2023jcp}. The contact maps of  Figure \ref{fig:fig6} report the (dimensionless) monomer distance $r_{ij}$ between monomers $i,j=1,\ldots,128$ for all monomers in the chain, and for increasing value of the bending rigidity $K$. They display characteristic patterns for the different structures. 
Without any bending energetic penalty ($K=0$) the chain collapses into a globule-crystal structure  as the temperature decreases  (see Figure \ref{fig:fig5}). Hence the corresponding contact map for $r_{ij}$ (Figure \ref{fig:fig6a}) does not display any significant structural features. 
Upon increasing $K$, the chain first achieves the twisted yarn-like structure displayed in Figure \ref{fig:fig6b}, and $r_{ij}$ acquires a well defined pattern in alternating blocks with high periodic high values occurring upon the $U-$ turns of the chains. This pattern continues and becomes more and more defined when switching to a rod-like structure, short (Figure \ref{fig:fig6c}) and long (Figure \ref{fig:fig6d}). We emphasize that, building on our previous discussion on the phase diagram of Figure \ref{fig:fig5}, all these three conformations belong to the same phase since their contact maps are very similar. A further increase beyond the critical value $K_c \approx 5$ leads to a toroidal structure with increasing internal radius, as illustrated in Figure \ref{fig:fig6e} for small toroid and in Figure \ref{fig:fig6f} for larger one. Here, the pattern of $r_{ij}$ is very different and presenting stripes associated with the spooling of the chain, with stripes becoming longer and less in number at increasing $K$ in view of the increasing radius of the toroid. Again, both the last two conformations pertain to the toroidal phase in our phase diagram of Fig. \ref{fig:fig5} that the contact maps recognize as different from the rod. These findings are fully consistent with the results of Ref. \cite{Aierken2023jcp}.
The difference between the two shapes can also be pinned down by using the eigenvalues of the moment of inertia, as explained in Supplementary Section S. 
This will be illustrated in some more compelling examples further below.

Having assessed the various phases in the phase diagram of Figure \ref{fig:fig5}, we now turn our attention to two additional interesting features of the phase diagram.

\begin{figure*}[htbp]
\centering
    \begin{subfigure}[b]{0.4\textwidth}
    \centering
    \includegraphics[width=9cm]{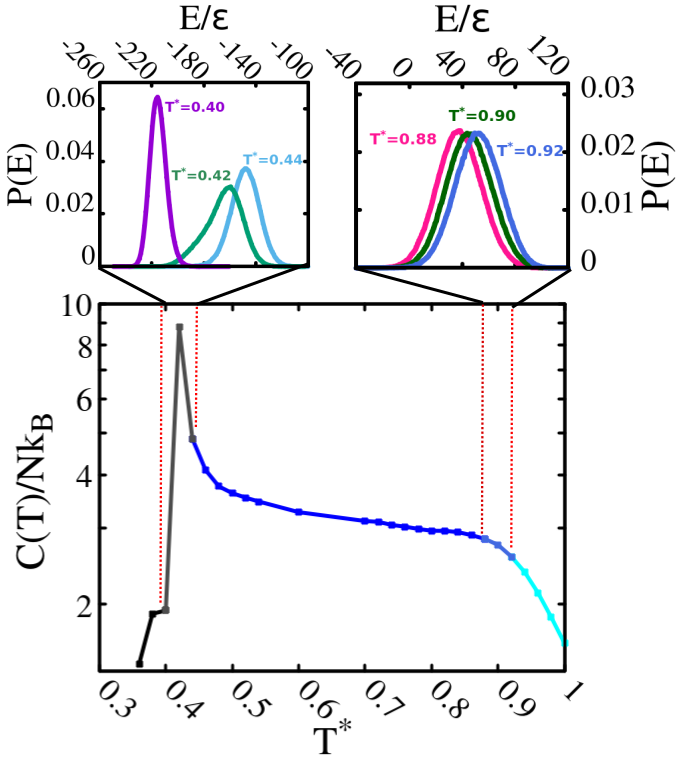}
     \caption{}\label{fig:fig7a}
    \end{subfigure}
    \qquad \qquad \qquad
    \begin{subfigure}[b]{0.4\textwidth}
    \centering
    \includegraphics[width=8cm]{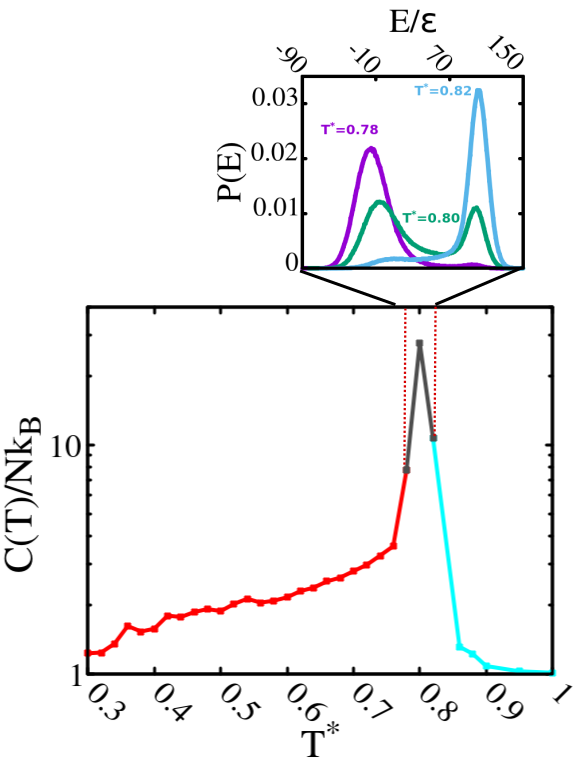}
     \caption{}\label{fig:fig7b}
    \end{subfigure}
\caption{Specific heat (dimensionless) per monomer as a function of the reduced temperature $T^{*}$ for different values of the bending stiffness $K$. (a) Case  $K=2$; (b) Case $K=9$. The insets depict the distribution $P(E)$ of the total potential energy $E$ at specific temperatures, with vertical dotted lines indicating the specific temperatures in the corresponding main figure.}
\label{fig:fig7}
\end{figure*}

The first one is related to the different character of the transition lines in the phase diagram of Figure \ref{fig:fig5}. Consider the coil-globule transition line that is only present in the interval $0 \le K \le K_c \approx 5$ and occurs in the approximate temperature range $0.8 \le T^{*} \le 0.9$. The crossing of the transition line  for $K=2$ is signaled by a kink in the reduced specific heat per monomer as indicated in Figure \ref{fig:fig7a} at $T^{*}\approx 0.90$ and the absence of a well pronounced peak is to be interpreted as a signature of a second-order transition as  happens also in the case of flexible polymer chain \cite{Taylor2009}. Here, a careful analysis \cite{Taylor2009} indeed confirms the existence of the transition even in the absence of a well defined peak in the specific heat. 
Upon further cooling, a marked peak in the specific heat is encountered at $T^{*} \approx 0.42$ indicating a first-order globule-crystal transition. 
Remarkably, a similar feature also occurs in a tangent bead model with side spheres \cite{Skrbic2019,Skrbic2019local,Skrbic2021marginally,Skrbic2021spontaneous}.
The different characters of the two transitions are also identified by the behaviour of the energy distribution $P(E)$ of the total potential energy (pair and bending) across the transition line. This is also shown in Figure \ref{fig:fig7a} as insets, where the $P(E)$ display a single peak which shifts to lower energies upon decreasing the temperature from $T^{*}=0.92$ to $T^{*}=0.90$ and to $T^{*}=0.88$ essentially unchanged in shape, whereas the same distribution presents a significant change in shape and in the location of the peak across the lower temperature transition from $T^{*}=0.44$ to $T^{*}=0.42$ and to $T^{*}=0.40$. Although not visible, we surmise the existence of a bimodal distribution for $P(E)$ exactly at the transition at $T^{*} \approx 0.42$, a signature of a first order transition. 
This bimodal distribution is indeed clearly visible in the single transition occurring at $K=9> K_c$ (Figure \ref{fig:fig7b}) where the marked peak in the specific heat occurring at $T^{*}=0.80$ is associated with a double peak at two different energies, whereas a single peak is visible at higher temperature $T^{*}=0.82$ and at lower temperature $T^{*}=0.78$.
Further evidence of the onset of a direct first order transition is reported in Supplementary Section S and Supplementary Figure SIV. 

The second point is even more interesting and it is related to the region labelled as \textbf{S} in the phase diagram of Figure \ref{fig:fig5}, which stands for ``Switching degenerate region'' which is also framing the putative tricritical point occurring when the second order coil-globule transition line meets the first-order coil-crystal line. Note that this point can also be regarded as a triple point where three different phases -- the coil, the globule, and the crystal, meet. This region extends vertically also across the rod-to-toroid transition when moving at higher $K$ at a fixed temperature $T^{*}$. What happens across this transition line and why does the conformation change from rod to toroidal structure in the first place? 
\begin{figure*}[htbp]
\centering
\begin{subfigure}[b]{0.4\textwidth}
    \centering
    \includegraphics[width=9cm]{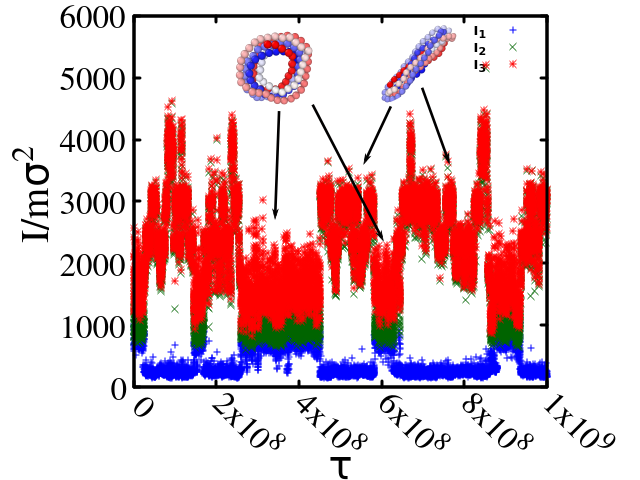}
     \caption{}\label{fig:fig8a}
    \end{subfigure}
    \qquad \qquad \qquad
    \begin{subfigure}[b]{0.4\textwidth}
    \centering
    \includegraphics[width=9cm]{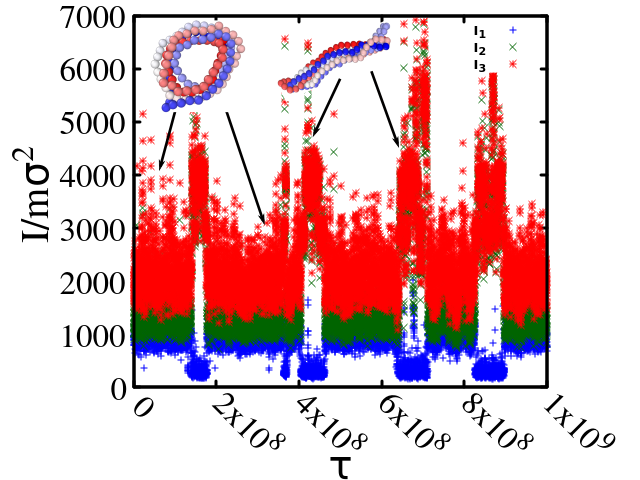}
     \caption{}\label{fig:fig8b}
    \end{subfigure}
    \qquad \qquad \qquad
    \begin{subfigure}[b]{0.4\textwidth}
    \centering
    \includegraphics[width=9cm]{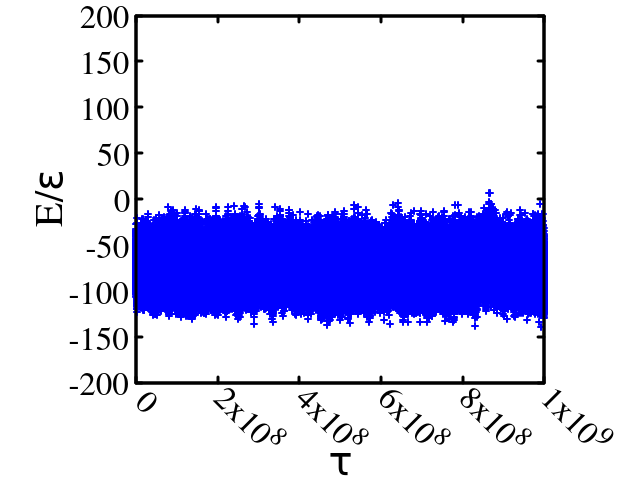}
     \caption{}\label{fig:fig8c}
    \end{subfigure}
    \qquad \qquad \qquad
    \begin{subfigure}[b]{0.4\textwidth}
    \centering
    \includegraphics[width=9cm]{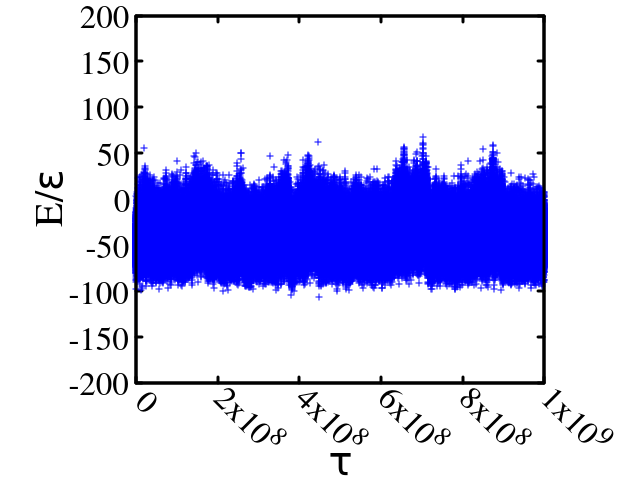}
     \caption{}\label{fig:fig8d}
    \end{subfigure}
\caption{Time evolution of the eigenvalues $I_1,I_2,I_3$ (ranked in ascending order) of the moment of inertia tensor $\mathbf{I}$ in reduced units $m \sigma^2$, as a function of time $\tau$, for (a) $K=5, T^*=0.62$ and (b) $K=6, T^*=0.70$. In both cases, the figure shows representative snapshots of the chain conformation that switches erratically from rod to toroid and viceversa. Panels (c) and (d) report the corresponding fluctuations in  total energy $E_{tot} = E_{bend} + E_{pair}$, with (c) corresponding to (a) and (d) corresponding to (b). 
The absence of any  significant variations of the average energy with respect to ordinary thermal fluctuations indicates that the two folded phases are isoenergetic in the two cases observed.}
\label{fig:fig8}
\end{figure*}
As anticipated in Section dedicated to order parameters \ref{subsec:order}, we can also distinguish between rod and toroidal structure also by the eigenvalues of their moment of inertia tensor $\mathbf{I}$. Both the rod and the toroid have uniaxial symmetry and hence they have two identical eigenvalues, but they can be distinguished because in the main eigenvector dictating their orientation in space is associated with the maximum eigenvalue in the former case and with the minimum eigenvalue in the latter case. Figure \ref{fig:fig8} shows the results for  the eigenvalues $I_1<I_2<I_3$ for $K=5$ and $T^{*}=0.62$ (Figure \ref{fig:fig8a}), that is a point at  the center of the \textbf{S} phase of the phase diagram in Figure \ref{fig:fig5}, and for $K=6$ and $T^{*}=0.70$ which lays on the right edge of the same phase. All three eigenvectors display significant fluctuations in their values, and this is especially true for the largest one $I_1$. What is interesting is the fact that in this \textbf{S} region the actual shape switches from rod to toroid very frequently, as indicated in the insets of both Figures \ref{fig:fig8a} and \ref{fig:fig8b}, indicating the degeneracy in energy between the rod and toroid in this region. See Figures \ref{fig:fig8c} and \ref{fig:fig8d}, which do not display any significant drift from the average total energy $E_{tot} = E_{bend} + E_{pair}$ during the corresponding swapping of the two different folds. This occurs because as $K$ increases, the bending energy $E_{bend}$ increases (see Figure \ref{fig:fig2a}) and hence the total energy $E$ also increases (Figure \ref{fig:fig2c}) and eventually the conformation switches from rod to toroid for any temperatures $T^{*}$ along the vertical transition line between the rod \textbf{R} and toroid \textbf{T} regions of the phase diagram \ref{fig:fig5}. In the switching region \textbf{S} this is particularly striking because of the significant thermal fluctuations occurring here. Again, this was noted before -- albeit not explicitly stated, in a previous recent study of the same model \cite{Aierken2023jcp} (see their Table I), as well as in another model in which the energy degeneracy is more significant \cite{Skrbic2019local,Skrbic2021marginally,Skrbic2021spontaneous}.
Additional evidence of this switching behaviour between isoenergetic structures within the switching region of the phase diagram \textbf{S} can be obtained by the analysis of the eigenvalues of the moment of inertia tensor, as reported in Supplementary Section S and Supplementary Figure SV and Supplementary Figure SVI. 
Note that a similar switching between rods of different lengths also occurs in the region of the phase diagram of Figure \ref{fig:fig5} between $K=4.2$ and $K=4.6$ and temperatures $T^{*}=0.52-0.54$ which lies at the edge of the switching region \textbf{S}, as explained in Supplementary Section S and visualized in Supplementary Figure SVII. 
\section{Results for multiple chains}
\label{sec:multiple}
\subsection{Self-assembly process}
\label{subsec:self}
Having assessed the phase behavior of the single semiflexible polymer chain, we now turn our attention to the self-assembly properties of  $N_c=80$ short ($N=12$) semiflexible polymer chains confined in a box at a given volume $V$ and a given temperature $T$. The system is equilibrated at a given temperature before collecting statistics. Inspired by past important work on the assembly of normally soluble proteins into amyloid fibrils \cite{Nguyen2004,Auer2007,Auer2008,Fuxreiter2021}, we consider different molar concentrations to identify the critical concentration sufficient to drive the nucleation process. This self-assembly process is expected to depend on temperature, concentration, and bending rigidity, and a guide to this large parameter space is provided by a recent mean field approach  \cite{Marcato2023} for semiflexible polymer chains on a lattice. While a gas-liquid transition was clearly identified (see their Figure 7), an expected isotropic-nematic transition was not observed in view of the polydispersity that is intrinsic to the method. This shortcoming will be resolved in the simulations of the present study. As further elaborated below, this isotropic-nematic transition is intertwined with the nucleation process occurring at low temperatures where all different polymer chains self-assemble into a fibril bundle where all chains tend to align along a common direction (due to their stiffness) and, at the same time, organize into a hexagonal packing. This process will be clearly identified by the relative value of the sum of the intra-chain interactions $E_{intra}$ (Eq. (\ref{sec1:eq8a})) and the sum of the inter-chain interaction $E_{inter}$ (Eq. (\ref{sec1:eq8b})), and it will be supported by the behavior of the isotropic-nematic order parameter $S$, as explained in Section dedicated to order parameters \ref{subsec:order}.
\begin{figure}[htbp]
\centering
\begin{subfigure}[b]{0.40\textwidth}
    \includegraphics[width=7cm]{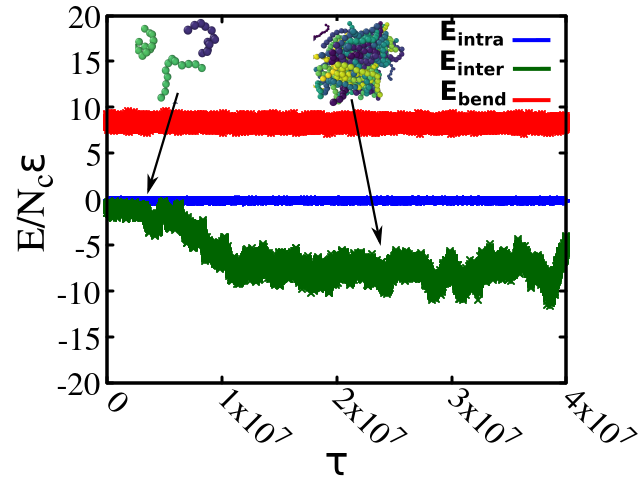}
     \caption{}\label{fig:fig9a}
    \end{subfigure}
    \begin{subfigure}[b]{0.40\textwidth}
    \includegraphics[width=7cm]{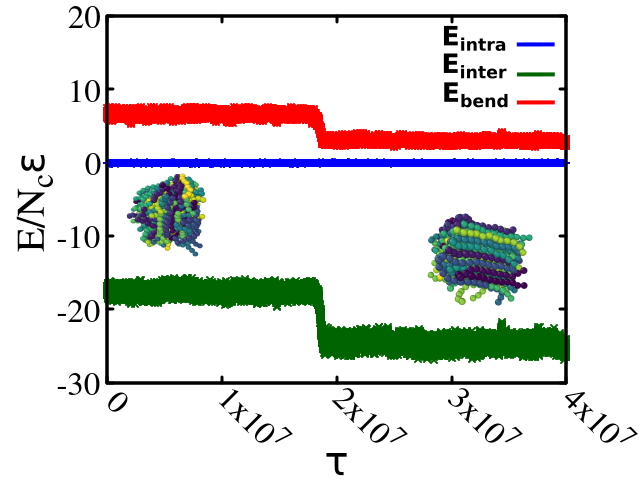}
     \caption{}\label{fig:fig9b}
    \end{subfigure}
    \begin{subfigure}[b]{0.40\textwidth}
    \includegraphics[width=7cm]{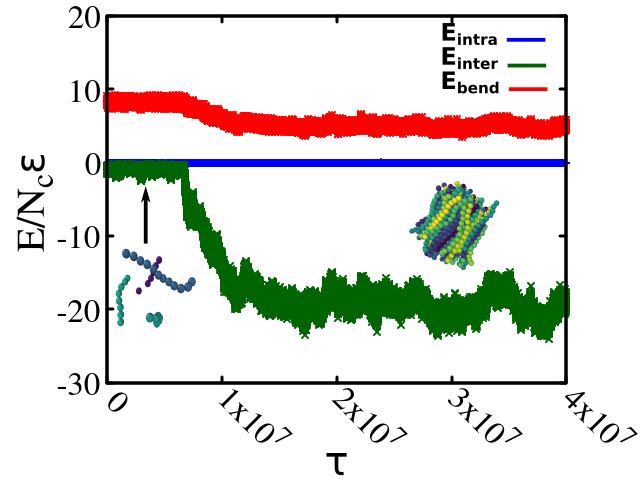}
     \caption{}\label{fig:fig9c}
    \end{subfigure}
\caption{Reduced energy components per \textcolor{black}{number of chains}  $E/N_c\epsilon$ as a function of the time and discriminated between $E_{inter}$ (Green crosses), $E_{intra}$ (Blue daggers), and $E_{bend}$ (Red filled squares)  \SI{10}{m\Molar{}} concentration and for different values of temperature $T^{*}$ and bending stiffness $K$. Insets depict representative snapshots at the specific time displayed.   (a) Close to the gas-globule transition for $K = 4$ and $T^* = 0.80$; (b) Close to the globule-fibril transition for $K = 4$ and $T^* = 0.68$;  (c) Close to the gas-fibril transition for $K = 8$ and $T^* = 0.82$. }
\label{fig:fig9}
\end{figure}
Figure \ref{fig:fig9} displays the reduced inter-chain energy per \textcolor{black}{chain} $E_{inter}/N_c\epsilon$ (Green crosses),  $E_{intra}/N_c\epsilon$ (Blue daggers), and $E_{bond}/N_c\epsilon$ (Red filled squares)  during the time evolution, for a system at \SI{10}{m\Molar{}} concentration and at different temperature $T^{*}$ and bending rigidity $K$. At $K = 4$ and $T^* = 0.80$ (Figure \ref{fig:fig9a}) $E_{inter}$ displays a drop after $\tau \approx 2.0 \times10^{7}$ time steps, a clear signature of a conformational change to a globule that maximize the favorable contacts, as also highlighted by the inset in Figure \ref{fig:fig9a}. 
During this process, neither $E_{intra}$ nor $E_{bend}$ display any significant deviation from their initial values, confirming that the aggregate is globule-like but formed by coil-like chains.

Consider now another state point with the same bending stiffness $K=4$ but at lower temperature $T^{*}=0.68$ (Figure \ref{fig:fig9b}). Because of the lower temperature compared with previous case, the magnitude of the various energy components is lower by a factor $2-3$. The snapshots displayed as insets in Figure \ref{fig:fig9b} indicate the system to be initially in the form of a gas of weakly interacting swollen chains that eventually tend to aggregate in a hexagonal bundle. Accordingly, $E_{inter}$ displays a marked drop at the corresponding time steps ($\tau \approx 2.0 \times10^{7}$) indicating that this nematic assembly compares energetically favourably with the globule one as it decreases the bending penalty, as indicated by the corresponding drop in  $E_{bend}$. As in the previous case, $E_{intra}$ does not display any significant change as expected for the assembly of swollen chains.  A combination of the results from Figures \ref{fig:fig9a} and \ref{fig:fig9b} shows that upon cooling at this mild stiffness of $K=4$, the chains first tend to assemble into a globular shape still maintaining their initial random swollen conformations, and then tend to form a hexagonal bundle via stiffening of the assembled chains upon further cooling.
If the coil state is the analog of the "gas" phase, the globule is the "liquid" phase, and the hexagonal bundle the "solid phase", in a process analogous to a gas-to-liquid-to-solid transition.
 
Next, we move along the $K$ axis in the phase diagram and consider a state point with a significantly higher value of $K=8$, and at a temperature $T^{*}=0.82$ comparable to the previous one. This is reported in Figure \ref{fig:fig9c} and displays a drastic drop in both $E_{inter}$ and $E_{bend}$ indicating the formation of a hexagonal bundle starting from a gas of weakly interacting chains, as indicated by the representative snapshots reported as insets, thus suggesting the presence of the analog of a direct gas-solid transition for this state point. Here, it is interesting to remark that this state point of $K=8$ and $T^{*}=0.82$ is also located close to the coil-toroid transition for the single chain (Figure \ref{fig:fig5}), thus still indicating that the nucleation effect from the chain ensemble prevails in the self-aggregation properties of the single chain (remember, however, that the length of the chains are different in this case). Here too, as in the previous two cases, $E_{intra}$ seems to be unaffected by the self-assembly process, as it keeps fluctuating around a constant value.

The formation of the hexagonal bundle corresponds to a isotropic-nematic transition that is further signaled by an abrupt increase of the nematic order parameter $S$ (see Supplemen- tary Section S and Supplementary Figure SVIII), 
and it can occur from the globule state (Figure \ref{fig:fig9b}) or directly from the coil state (Figure \ref{fig:fig9c})

\subsection{Free energy landscape}
\label{subsec:free}

\begin{figure}[htbp]
\centering
\begin{subfigure}[b]{0.40\textwidth}
    \centering
    \includegraphics[width=7cm]{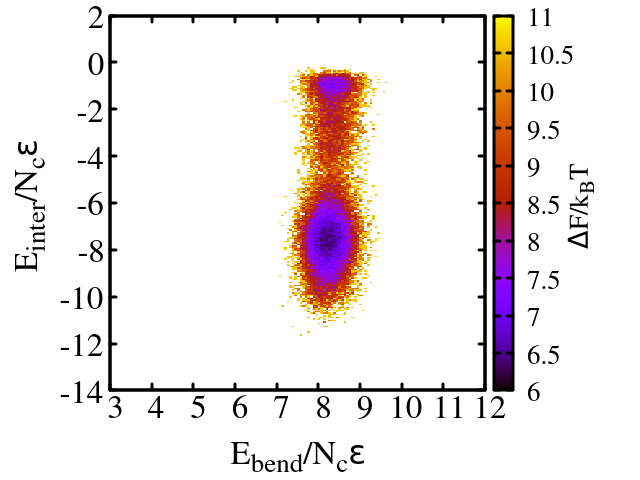}
     \caption{}\label{fig:fig10a}
    \end{subfigure}
     \qquad \qquad \qquad
    \begin{subfigure}[b]{0.40\textwidth}
    \centering
    \includegraphics[width=7cm]{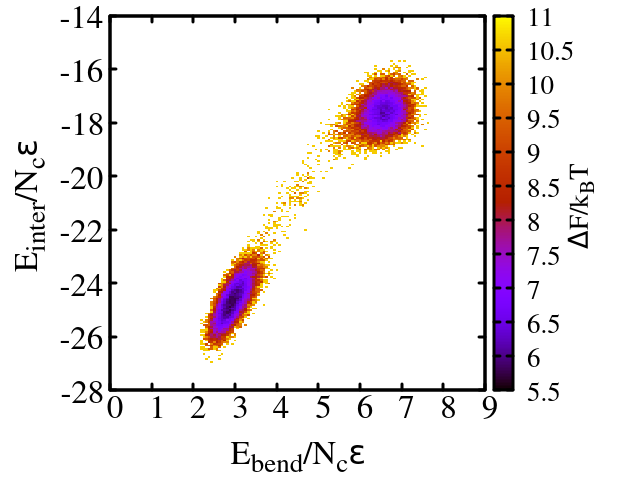}
     \caption{}\label{fig:fig10b}
    \end{subfigure}
    \qquad \qquad \qquad
     \begin{subfigure}[b]{0.40\textwidth}
    \centering
    \includegraphics[width=7cm]{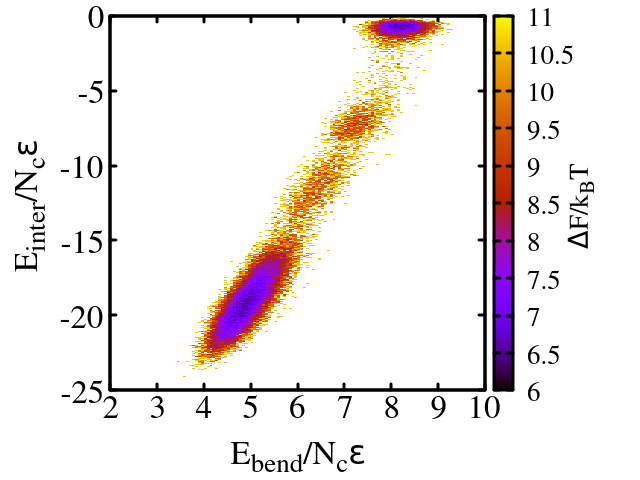}
     \caption{}\label{fig:fig10c}
    \end{subfigure}
\caption{Free energy landscape analysis in terms of $E_{inter}$ and $E_{bend}$ at the same state points of Figure \ref{fig:fig9} and at the same  \SI{10}{m\Molar{}}) concentration   (a) Close to the coil-globule transition for $K = 4$ and $T^* = 0.80$; (b) Close to the globule-bundle transition for $K = 4$ and $T^* = 0.68$; (c) Close to the direct coil-bundle transition for $K = 8$ and $T^* = 0.82$. The right colored bar represents the change in free energy in unit of thermal energy $\Delta F/k_B T$ color coded from high (light color) to low (dark color) changes.}
\label{fig:fig10}
\end{figure}
Our findings reported in Figure \ref{fig:fig9} are suggestive of a nucleation process leading to the formation of hexagonal bundles. This suggestion is further supported by a free energy landscape using $E_{inter}$ and $E_{bend}$ as reaction coordinates. This is reported in the contour plots of Figure \ref{fig:fig10} for the same state points of Figure \ref{fig:fig9}. In the case of the coil-globule transition, reported in Figure \ref{fig:fig10a}, the two minima are nearly vertically aligned at constant $E_{bend}$ and are associate with the coil (top) and globular (bottom) conformations of Figure \ref{fig:fig9a}. Close to the globule-bundle transition at $K = 4$ and $T^* = 0.68$ (Figure \ref{fig:fig10b}), we observe two deep minima corresponding to the two main conformations (globule and bundle) displayed in Figure \ref{fig:fig9b}.  The final energy landscape of Figure \ref{fig:fig10c} corresponds to the direct coil-bundle transition of Figure \ref{fig:fig9c} and display two well defined minima corresponding to the gas phase of swollen chains (top right) and to the hexagonal bundle formation (bottom left).

\subsection{Phase diagram for multichains}
\label{subsec:phase_multi}
All the results of previous two Sections \ref{subsec:self} and \ref{subsec:free} were derived for $N_c=80$ semiflexible polymer chains formed by $N=12$ monomers at concentration of \SI{10}{m\Molar{}})  or volume fraction $\phi = 3.36 \times 10^{-3}$. This is a rather low concentration that was selected to match the value used by past similar studies for different models \cite{Nguyen2004,Auer2007,Auer2008}. Nguyen \textit{et al} \cite{Nguyen2004} used a coarse-grained version of short peptides in implicit solvents, and studied the self-assembly process for concentrations ranging from \SI{0.5}{m\Molar{}}) to \SI{20}{m\Molar{}}); Auer \textit{et al} \cite{Auer2007} used a simplified version of a 'thick-polymer'  model \cite{Hoang2004} again in a concentration regime \SI{1}{m\Molar{}} to \SI{20}{m\Molar{}}. In both cases, the self-assembly process leading to the formation of fibril bundles was studied in some detail, but none of them reported the corresponding phase diagram. On the other hand, a recent mean field theory based on a spin-model on a lattice \cite{Marcato2023} underscored the presence of a gas-liquid transition but failed to identified additional transitions associated with the bending rigidity $K$ that is expected to play a significant role \cite{Ivanov2003}, in view of the strong polydispersity in the lengths of the chains that was intrinsic in the mean field approach. 

\begin{figure*}[htbp]
\centering
\begin{subfigure}[b]{1.0\textwidth}
    \centering
    \includegraphics[width=9cm]{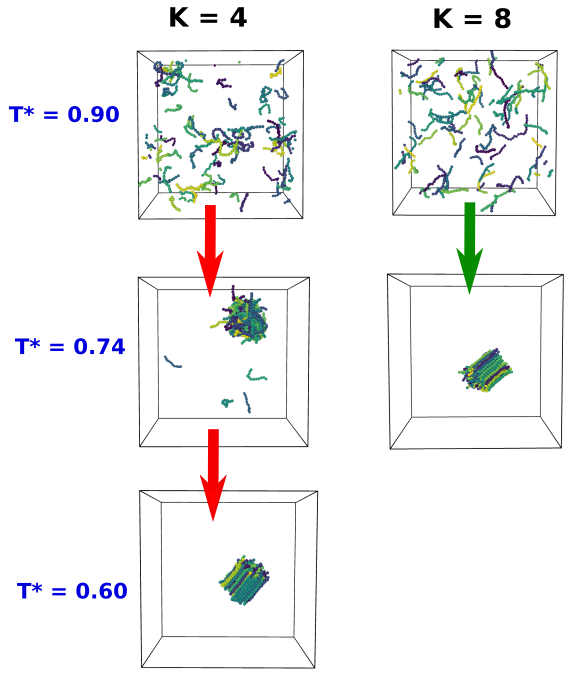}
     \caption{}\label{fig:fig11a}
    \end{subfigure}
     \qquad \qquad \qquad
    \begin{subfigure}[b]{1.0\textwidth}
    \centering
    \includegraphics[width=9cm]{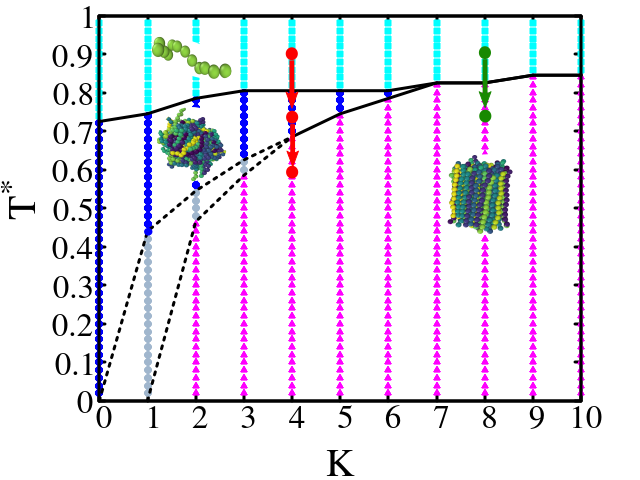}
     \caption{}\label{fig:fig11b}
    \end{subfigure}
\caption{Self-assembly of a system of  $N_c=80$ semiflexible polymer chains formed by $N=12$ monomers at concentration of \SI{10}{m\Molar{}}  or volume fraction $\phi = 3.36 \times 10^{-3}$.  (a) The two pathways followed by the nucleation process depending on the value of $K$. For $K=4$ chains are sufficiently flexible to first nucleate a globular structure G coexisting with free chains, and then they form a bundle B upon further cooling through an ordering process (red path). For $K=8$ the two processes are combined (green path). (b) Phase diagram in the temperature $T^{*}$-bending rigidity $K$ plane for the same system highlighting the two paths described in (a). }
\label{fig:fig11}
\end{figure*}

We are now in the position of extending the temperature $T^{*}$-bending rigidity $K$ single chain phase diagram of Figure \ref{fig:fig5} to multichains and, at the same time, making the connection with the aforementioned studies. Figure \ref{fig:fig11} reports the extension of the single chain phase diagram of Figure \ref{fig:fig5} to many chains for a \SI{10}{m\Molar{}} molar concentration  or for a volume fraction $\phi = 3.36 \times 10^{-3}$, obtained using the results of previous Sections \ref{subsec:self}  \ref{subsec:free}. Figure \ref{fig:fig11a} shows two different pathways followed upon cooling depending on the value of the bending rigidity $K$. For small $K$ (for instance $K=4$ in Figure \ref{fig:fig11a}) different semiflexible chains first form a globular assembly coexisting with some free chains through a nucleation process, and then order  by aligning the chains into a hexagonal lattice (nucleation $\rightarrow$ ordering, red path). For larger $K$ ($K=8$ in the present case) the two process occurs simultaneously (nucleation+ordering, green path) and there is a direct formation of the hexagonal bundle. Interestingly, this is exactly the situation in Ref. \cite{Auer2007,Auer2008} notwithstanding the different model used in that study, the only difference being the final self-assembled structure, a fibril in that case, a hexagonal bundle in the present study. 

Armed by these findings, we can now draw the phase diagram of the self-assembly process in the temperature $T^{*}$-bending rigidity $K$ plane, as we did for the single chain counterpart (Figure \ref{fig:fig5}). Consider for instance the case $K=4$ discussed in Figures \ref{fig:fig9} and \ref{fig:fig10}. On cooling from a high temperature gas of swollen chains, one first encounters a coil-globule transition close to $T^{*}=0.80$, and then a globule-bundle transition close to $T^{*}=0.68$. In both cases the self-assembly process is signaled by the behavior of the energy components, as previously indicated in Figures \ref{fig:fig9} and \ref{fig:fig10} and by an upswing of the nematic order parameter $S$ (see Supplementary). As $K$ increases, the extension of the globular region gradually decreases until it eventually disappears. This was indeed indicated by the results of Figures \ref{fig:fig9c} and \ref{fig:fig10c} where for $K=8$ and $T^{*}=0.82$, that is well above to the triple point occurring at $K_c \approx 7$ and $T_c^{*} \approx 0.8$ of Figure \ref{fig:fig11} but still close to the coil-bundle transition. The $N_c$ semiflexible chains self-assemble directly into a bundle, without any intermediate, and this is the counterpart of the companion transition reported in Figure \ref{fig:fig5}.

The above coil-bundle transition is the analog of the first order isotropic-nematic transition observed by Ivanov \textit{et al} \cite{Ivanov2003} using grand-canonical Monte Carlo simulations of semiflexible polymer chains of $20$ monomers (see their Figure 7) and clearly occurs only at sufficiently high concentrations. Note however that 
the present \SI{10}{m\Molar{}}) molar concentration  corresponding to a volume fraction $\phi = 3.36 \times 10^{-3}$, is significantly smaller than the typical concentration of attractive rods or slender helices with similar aspect ratios \cite{DalCompare2023}.
We do expect, however, no self-assembly below a certain concentration that depends on the temperature.

More generally, one could study how the phase diagram presented in Figure \ref{fig:fig11b} would change with concentration, thus including an additional concentration axis to it. This would allow the observation of the full phase diagram suggested  by a recent mean field theory \cite{Marcato2023} and already observed experimentally for flexible polymers \cite{Olmsted1998} and partially explored numerically \cite{Sheng1994,Sheng1996}. Consider the flexible ($K=0$) case first. Upon increasing the volume fraction, a coil-globule binodal (akin to a liquid-liquid binodal) is expected first, followed by a globule-bundle transition \cite{Olmsted1998}. The presence of a bending rigidity ($K>0$) is likely to shrink the coil-globule binodal, in view of the results of Figure \ref{fig:fig11}, until eventually a direct crystallization occurs.

By decreasing the concentration to \SI{1}{m\Molar{}}) molar concentration (Supplementary materials), we find all transition lines reported in Figure \ref{fig:fig11b} are shifted to lower temperatures, until eventually they disappear below some critical concentration. This is indeed the signature of the gas-liquid transition predicted by a mean field theory on a lattice \cite{Marcato2023}. Note that, however, a detailed characterization of this process  would require a dedicated (grand-canonical) calculation.

\begin{figure}
    \centering
    \includegraphics[width=8cm]{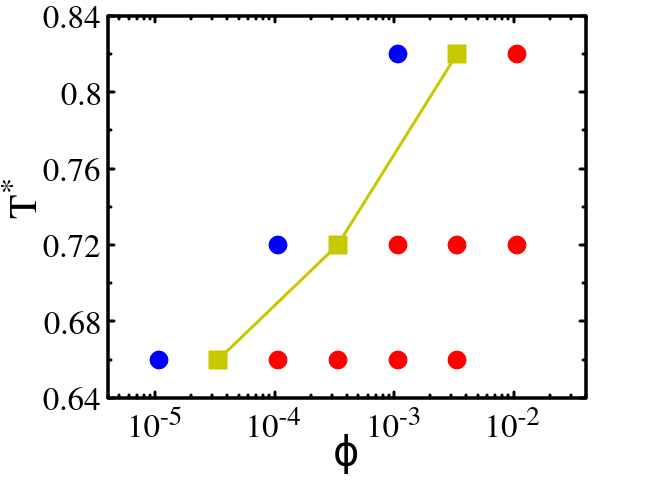}
    \caption{Reduced temperature $T^{*}$ as a function of the concentration reporting the critical temperature for nucleation at different volume fractions $\phi$ and at a fixed value of $K=8$. For any fixed temperature, the system is in the coil phase (blue filled circles) until a critical concentration is reached (yellow solid squares). At higher concentrations (red solid circles) the system is in a hexagonal bundle phase}
    \label{fig:fig12}
\end{figure}

The way in which the critical nucleation concentration changes with temperature is displayed in Figure \ref{fig:fig12} which reports the two different conformations, free coil chains (solid blue circles) and hexagonal bundles (solid red circles) obtained below and above the critical concentration at that given temperature. Here a bending rigidity of $K=8$ has been used which, in the phase diagram of Figure \ref{fig:fig11b}, corresponds to a direct transition from coil to bundle. Hence, this is the case where a direct crystallization of the polymer chains preempts the liquid-liquid transition that becomes metastable \cite{Olmsted1998}. 
For any given reduced temperature $T^{*}$, at very low $\phi$ all chains are free and hence all chains are in the form of free coils. For all displayed temperatures, however, one observes nucleation to a hexagonal bundle at a given concentration (green filled squares) which then remains the stable structure at any higher concentration (red filled circles). 

Note that Figure \ref{fig:fig12} is reminiscent of the calculation of the critical micelles concentration that is usually found in the self-assembly of surfactants \cite{Carrer2020} and Janus-like systems \cite{Sciortino2009,Sciortino2010}. However, it should be emphasized that the mechanism presented here is a phase separation at a critical concentration rather than micellization where a main cluster coexists with monomers. 

In closing this Section, it is important to remark that we have considered here only low density regimes where all complications stemming from topological interactions characteristic of polymer melts at high density are nevertheless observed \cite{Auhl2003,Shakirov2018,Bobbili2020,Kawak2021,Kos2021}.

\section{Conclusions}\label{sec:conclusions}
The motivation of the present study stems from a recent important work by \citeauthor{Fuxreiter2021} \cite{Fuxreiter2021} who discussed the nature of the condensed states in proteins aggregation as analogs of the conventional states of matter, evaluating inter-chain and intra-chain interactions. When intra-chain interactions are dominant with respect to inter-chain interactions, then each protein folds independently by reaching its own native state. This is the analog of a 'gas phase'. In the opposite regime when inter-chain interactions dominate the intra-chain counterpart, different proteins tend to align to form a bundle via a nucleation process, thus forming an amyloid state. This is the analog of a 'solid state'. These authors then identified the intermediate droplet state as the counterpart of the liquid state in conventional condensed matter. This can be reached by the proteins in their native state via liquid-liquid phase separation \cite{Hansen2023}, and then progress to the amyloid state through a maturation process.

This nucleation process was already studied by Auer \textit{et al} \cite{Auer2007,Auer2008} using a model different from a semiflexible chain that was devised specific for describing peptides \cite{Hoang2004}. A similar calculation with a rather different model was performed by Nguyen \textit{et al} \cite{Nguyen2004} thus pointing toward a general self-assembly mechanics largely independent on the behavior of the single chain.

While it is now well established that a semiflexible polymer chain cannot accurately represent a protein, it shares with the protein  the notion of a well defined 'native state', unlike the flexible polymer chain whose ground state is a highly degenerate structureless globule. In addition, liquid-liquid phase separation also occurs for DNA droplets \cite{Wilken2023}, where the semiflexible polymer, as potentially represented by the present model, is indeed relevant. 

Using Langevin Dynamics simulations in implicit solvent, we have studied its phase diagram in the temperature $T^{*}$-bending $K$ rigidity plane. We found two different low temperature regimes for small and large bending rigidity. For $K<K_c\approx 5$ on cooling we first observed a second order coil-globule transition, followed by a first order globule-crystal transition upon further temperature decrease. As $K$ increases, the extension of the intermediate globular region decreases until eventually it disappears when $K=K_c \approx 5$, so that   a direct coil-to-crystal transition is observed above this value. The observed low-temperature states also depend on the value of $K$ (see Figure \ref{fig:fig5}). In the case of fully flexible polymer chain ($K=0$) a FCC-like crystal structure is observed in agreement with past extensive work on this system \cite{Taylor2009}. For $0<K<K_c \approx 5$ the crystal has a rod and twisted shape that occurs because it maximize the number of favourable contacts with bending energy penalties only at the edges. Above $K_c=5$, the bending penalties at the edges become so strong that the chain prefers to distribute it through a twisted toroidal shape which still preserving a sufficiently large number of favorable contacts. While other possible shapes (knotted, hairpins, double helices) have been observed in the past as low temperature states \cite{Montesi2004,Seaton2013,Marenz2016,Majumder2021,Du2022,Aierken2023pre}, evidence of the stability of these additional intermediate structures is unclear, whereas the stability of rods and toroids is supported by theoretical arguments \cite{Ishimoto2006,Hoang2014,Hoang2015}, numerical simulations \cite{Pham2005,Lappala2013,Kong2014,Hoang2014,Hoang2015,Dey2017,Wu2018role,Nguyen2023}, and by experimental evidence in DNA condensation \cite{Vilfan2006}. Our single chain phase diagram also confirms on-lattice Monte Carlo simulation results \cite{Bastolla1997,Doye1998} supported by mean-field theory \cite{Doniach1996,Lise1998} that derived a qualitatively similar phase diagram that differs from the present one by the presence of just a single (rod-like) low temperature phase, the only one compatible with the lattice structure.
It is also worthwhile to emphasize that the point $K_c \approx 5$ and $T_c^{*} \approx 0.80$ is \textit{both} a triple point, where the coil, globule and crystal phase meet, \textit{and} a tricritical point, where a second and first order line meet.
The final very interesting feature of our single chain phase diagram is the presence of a switching phase framing the triple point from below. Here, rods and toroids with specific shapes have nearly identical energies and they are observed to switch conformation from one to another driven by thermal fluctuations. This was not noted before in this framework, but it is reminiscent of a similar phenomenon occurring in a model for a flexible polymer chain with side spheres \cite{Skrbic2019local,Skrbic2021spontaneous}.
\textcolor{black}{A fully-fledged knowledge of the single chain phase diagram is of considerable importance in interpreting force-extension measurements in nucleic acids \cite{Marko1995,Bouchiat1999}, as it can guide the experiments in probing the most interesting region of the phase diagram. }

By dispersing several such semiflexible polymer chains in a box at a given volume and temperature, we then studied the extension of the single chain phase diagram to multichains. In this case, a gas of such chains in a swollen disordered coil state is observed to self-assemble into a globular structure at low bending stiffness, and into a nematic bundle where different chains align along a common direction to form a hexagonal lattice in the perpendicular direction. Upon lowering the concentration, the location of the transition lines progressively  shift to lower temperatures, until eventually disappearing at the critical point. However, to be able to see this gas-liquid transition, that has been predicted by mean field arguments \cite{Marcato2023}, a grand-canonical approach is required. On the other hand, the multichain phase diagram of Figure \ref{fig:fig11} predicts the existence of an isotropic-nematic transition that was not observed in mean field theory because of intrinsic polydispersity present in the method. 

The self-assembly process observed in the present study for semiflexible polymer chains at sufficiently low temperatures is also reminiscent of a similar nucleation process observed in the past with other models \cite{Ivanov2003,Nguyen2004,Auer2007,Auer2008} and it might stimulate further activities in the understanding of the putative liquid-liquid phase separation phenomena occurring in proteins aggregation. It would also be interesting to study phase separation in mixtures of flexible and semiflexible polymers \cite{Adhikari2011,Park2020} using the approach presented in this study.
\textcolor{black}{Another case in which the findings of the present study may be relevant is the organization of the bacteriophage dsDNA genome into the capsids that can be monitored via cryo-electron microscopy \cite{Leforestier2010}.}

\begin{acknowledgement}
The uses of the SCSCF and vHPC multiprocessor clusters at  the Universit\`{a} Ca' Foscari Venezia are gratefully acknowledged. We acknowledge the CINECA
awards HP10CGFUDT, HP10C1XOOJ, HP10CEB73V, HP10C7XMSY and HP10CG92F4 for the availability of high-performance computing resources and support under the ISCRA initiative. The work was supported by MIUR PRIN-COFIN2022 grant 2022JWAF7Y (AG). 
\end{acknowledgement}

\appendix
\section{Supporting Information}
\subsection{Moments of Inertia of toroids and rods}
\label{subsec:moments}

\begin{figure}[h]
    \centering
    \includegraphics[trim=0 0 0 0,clip,width=0.5\linewidth]{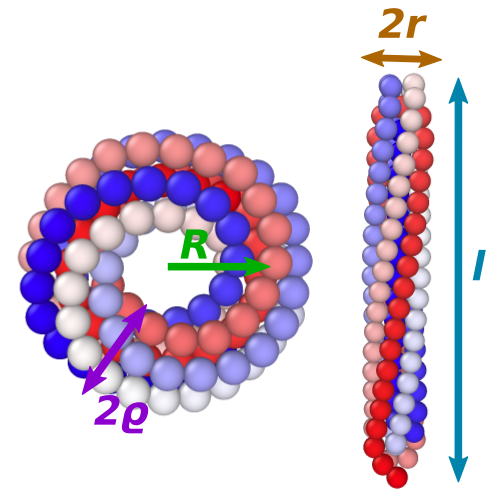}
    \caption{Characteristic geometrical parameters associated with the toroidal and rod geometries.}
    \label{fig:figs1}
\end{figure}

The time trajectories of $I_1(\tau),I_2(\tau),I_3(\tau)$ are obtained via diagonalization of the inertia tensor defined by:
\begin{equation}
I=m\sum_i
\begin{bmatrix}
{y_i}^2+{z_i}^2 & -{x_i}{y_i} & -{x_i}{z_i}\\
-{x_i}{y_i} & {x_i}^2+{z_i}^2 & -{y_i}{z_i}\\
-{x_i}{z_i} & -{y_i}{z_i} & {x_i}^2+{y_i}^2
\end{bmatrix}
\label{matr_ine}
\end{equation}
where the entries of Eq.(\ref{matr_ine}) are computed every timestep $\tau$ using the instantaneous beads positions $x_i,y_i,z_i$ measured with respect to the center of mass of the system.
The beads mass is always assumed to be unity, i.e. $m=1$.
The resulting triplet of eigenvalues is sorted in ascending order with $I_1<I_2=I_3$.
To show how the above defined quantities are related to the conformation topology, we consider simple geometries.
Modelling the rod as a cylinder of length $l$ and base radius $r$ (Figure \ref{fig:figs1}), the diagonalized inertia tensor is immediately given by: 
\begin{equation}
I_{rod}=
\begin{bmatrix}
\frac{1}{12}m\left(3r^2+l^2\right) & 0 & 0\\
0 & \frac{1}{12}m\left(3r^2+l^2\right) & 0\\
0 & 0 & \frac{1}{2}mr^2
\end{bmatrix}
\label{ine_rod}
\end{equation} 
Under the condition $l>\sqrt{3}r$, verified for all the rods observed in the simulation, the axial eigenvalue $I_{1,rod} = \frac{1}{2}mr^2$ represents the smallest element in the triplet.
The other two degenerate eigenvalues $I_{2,rod} = I_{3,rod} = \frac{1}{12}m\left(3r^2+l^2\right)$ have a quadratic dependence on the rods length.
This property justifies the use of $I_{2,rod}$ (or $I_{3,rod}$) as order parameters to distinguish rods of different lengths in the coexistence regimes, as reported in Figure \ref{fig:figs5}.
The toroid shape can be roughly approximated by a torus of major radius $R$ and minor radius $\varrho$.
The moment of inertia tensor  for a torus geometry is then given by the diagonal tensor
\begin{equation}
I_{tor}=
\begin{bmatrix}
\frac{1}{8}m\left(5\varrho^2+4R^2\right) & 0 & 0\\
0 & \frac{1}{8}m\left(5{\varrho}^2+4R^2\right) & 0\\
0 & 0 & \frac{1}{4}m\left(3{\varrho}^2+4R^2\right)
\end{bmatrix}
\label{ine_rod2}
\end{equation}

Unlike the previous case where $I_{1,rod} \ll I_{2,rod}=I_{3,rod} $, the eigenvalues $I_{1,tor},I_{2,tor},I_{3,tor}$ are quantities of the same order of magnitude due to the presence, in the analytical expressions, of both $R$ and $\varrho$.
The strategy adopted to effectively distinguish the rod configuration from coexisting toroids starts from the preliminary observation that $\varrho\approx r$ and the characteristic dimensions $R$ and $l$ are in general much bigger than the section radii: $R \gg \varrho,r$ , $l \gg \varrho,r$ (Figure \ref{fig:figs1}).
Hence, by comparing the rod eigenvalue $I_{1,rod} = \frac{1}{2}mr^2$, dependent only on $r$, with the three eigenvalues $I_{1,tor},I_{2,tor},I_{3,tor}$ found for the torus, it is possible to deduce that $I_{1,rod} \ll I_{1,tor},I_{2,tor},I_{3,tor}$ and, therefore, $I_{1,rod} \ll I_{1,tor}$.
$I_1$ represent, therefore, a good order parameter to identify and distinguish toroid and rods configurations. It is worth noticing that the independence of $I_{1,rod}$ on the rod length $l$ assures that any jump observed in $I_1(\tau)$ can be uniquely associated to a rod-toroid transition.
The reliability of the method discussed above can be successfully verified in the trajectory reported in FIG SV where toroid-rod transitions are accompanied to a visible jump of the quantity $I_1(\tau)$. In contrast, fluctuations in the rod length don not affect the trajectory of $I_1$ which remains constant on average while $I_2$ and $I_3$ follow the variations of $l$.

\subsection{Volume fractions}
\label{subsec:volume}
\begin{figure}[h]
    \centering
    \includegraphics[trim=0 0 0 0,clip,width=0.5\linewidth]{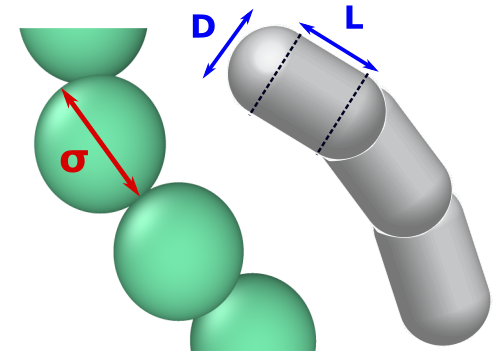}
    \caption{Spherical beads chain model (in green on the left) and spherocylindrical beads chain model (in gray on the right) depicting the corresponding characteristic geometrical parameters.}
    \label{fig:figs2}
\end{figure}

To reproduce the same volume fraction used in the simulations of Auer et al \cite{Auer2007}, we firstly derive the size of the simulation box and the portion of it occupied by a gas of $N_c = 80 $ and molar concentration $C$.
For a tubular beads polymer comprised of 11 jointed spherocylindrical segments of length $L$ and section diameter $D$ , the volume occupied by a single polymer chain is estimated as: \[V_{tub}= {11L\frac{\pi}{4}D^2 + \frac{\pi}{6}}D^3\] where the second term of the sum takes into account the two hemispheres at the ends of the chain.
For a known molar concentration $C$, the size of the simulation box used in Auer et al. model can be computed as \[V_{box}= {\frac{N_c}{N_A C}}\]
where $N_A$ is the Avogadro number.
The volume fraction is, therefore, obtained by the ratio: \[\phi=\frac{N_cV_{tub}}{V_{box}} = V_{tub} N_A C\]
In our simulations, the number of chains $N_c = 80$ is then kept constant while the simulation box is appropriately rescaled to match the target volume fraction $\phi$.
The latter operation requires an estimation of the volume occupied by the spherical beads polymer implemented in MD simulation.
By considering the polymer chain as a sequence of 12 tangent hard spheres of diameter $\sigma \approx 1$ , the volume occupied by a single chain is given by \[V_{sph} = 12V_{mon} = 2\pi{\sigma}^3\] hence the box volume for our molecular dynamics (MD) simulations is simply given by \[V_{box,MD}=\frac{N_cV_{sph}}{\phi}\]
Setting $D= 4 AA$ and $ L = 3.8 AA$ $C = 10 mM$, consistently with the parameters used by Auer and al, we obtain a volume fraction of $\phi = 3.36 \times 10^{-3}$.       
\subsection{Radial correlation function for single polymer chain}
\label{subsec:radial}
\begin{figure*}
    \centering
    \includegraphics[trim=0 0 0 0,clip,width=0.8\linewidth]{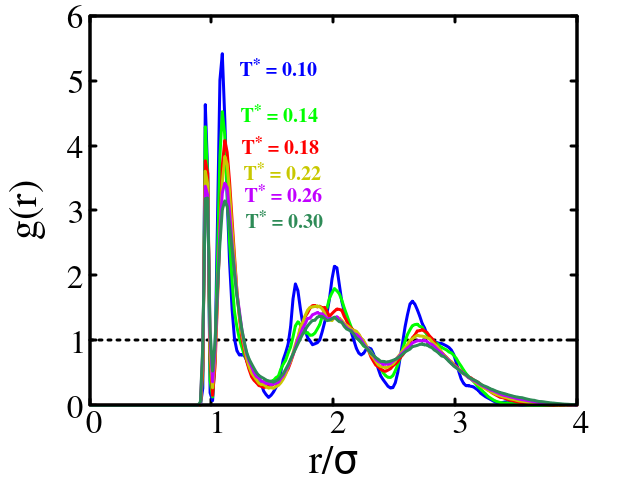}
    \caption{Radial distribution function $g(r)$ as a function of $r/\sigma$ at $K=0$ for reduced temperatures $T^*=0.10$, $T^*=0.14$, $T^*=0.18$, $T^*=0.22$, $T^*=0.26$,$T^*=0.30$.}
    \label{fig:figs3}
\end{figure*}

The comparison between the different curves in the radial distribution function $g(r)$ of Figure \ref{fig:figs3} shows the progressive disappearance of the extra peaks associated to the crystalline arrangement.
\subsection{Specific heat and energy probability distribution for single polymer chain}
\label{subsec:specific}
\begin{figure*}[h]
\centering
\begin{tikzpicture}
\node[anchor=south west,inner sep=0] (image) at (0,0){\scalebox {0.32}{\includegraphics[width=1\textwidth, trim=4cm 6cm 8cm 0cm, clip=true, angle=0, page=1]{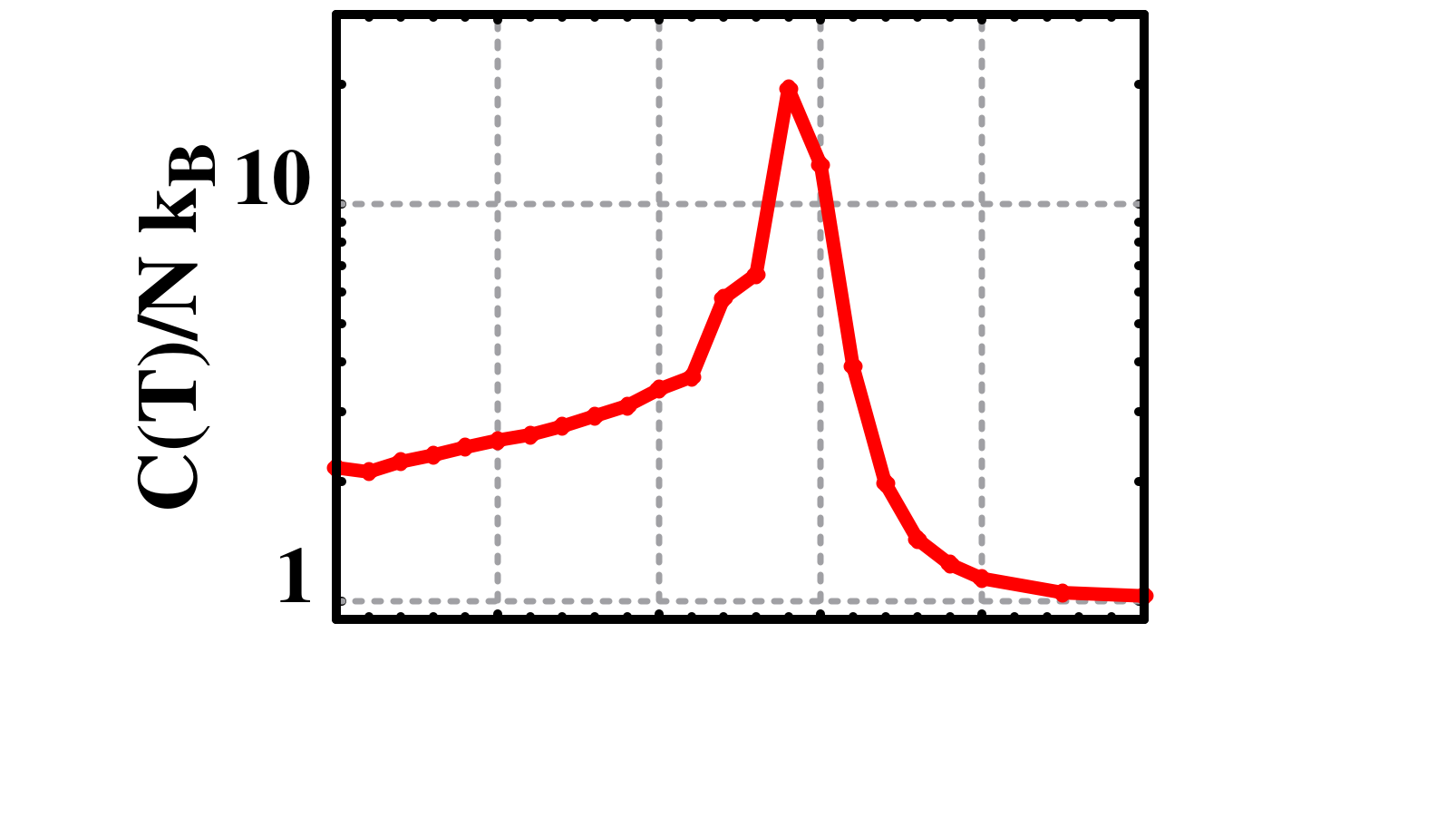}}};
\end{tikzpicture}
\begin{tikzpicture}
\node[anchor=south west,inner sep=0] (image) at (0,0){\scalebox {0.32}{\includegraphics[width=1\textwidth, trim=3.5cm 6cm 8cm 0cm, clip=true, angle=0, page=1]{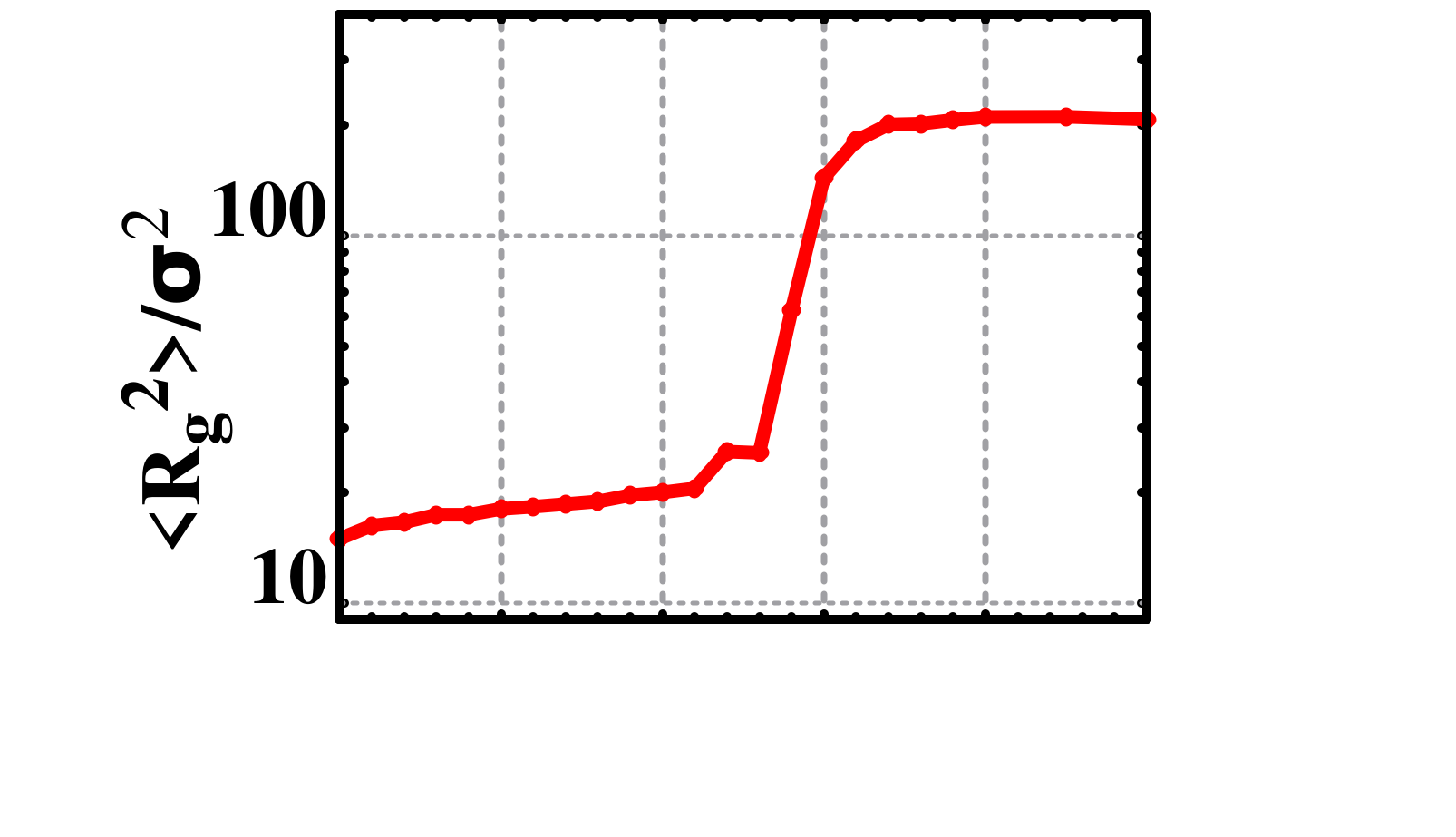}}};
\end{tikzpicture}
\begin{tikzpicture}
\node[anchor=south west,inner sep=0] (image) at (0,0){\scalebox {0.34}{\includegraphics[width=1\textwidth, trim=2cm 6cm 8cm 0cm, clip=true, angle=0, page=1]{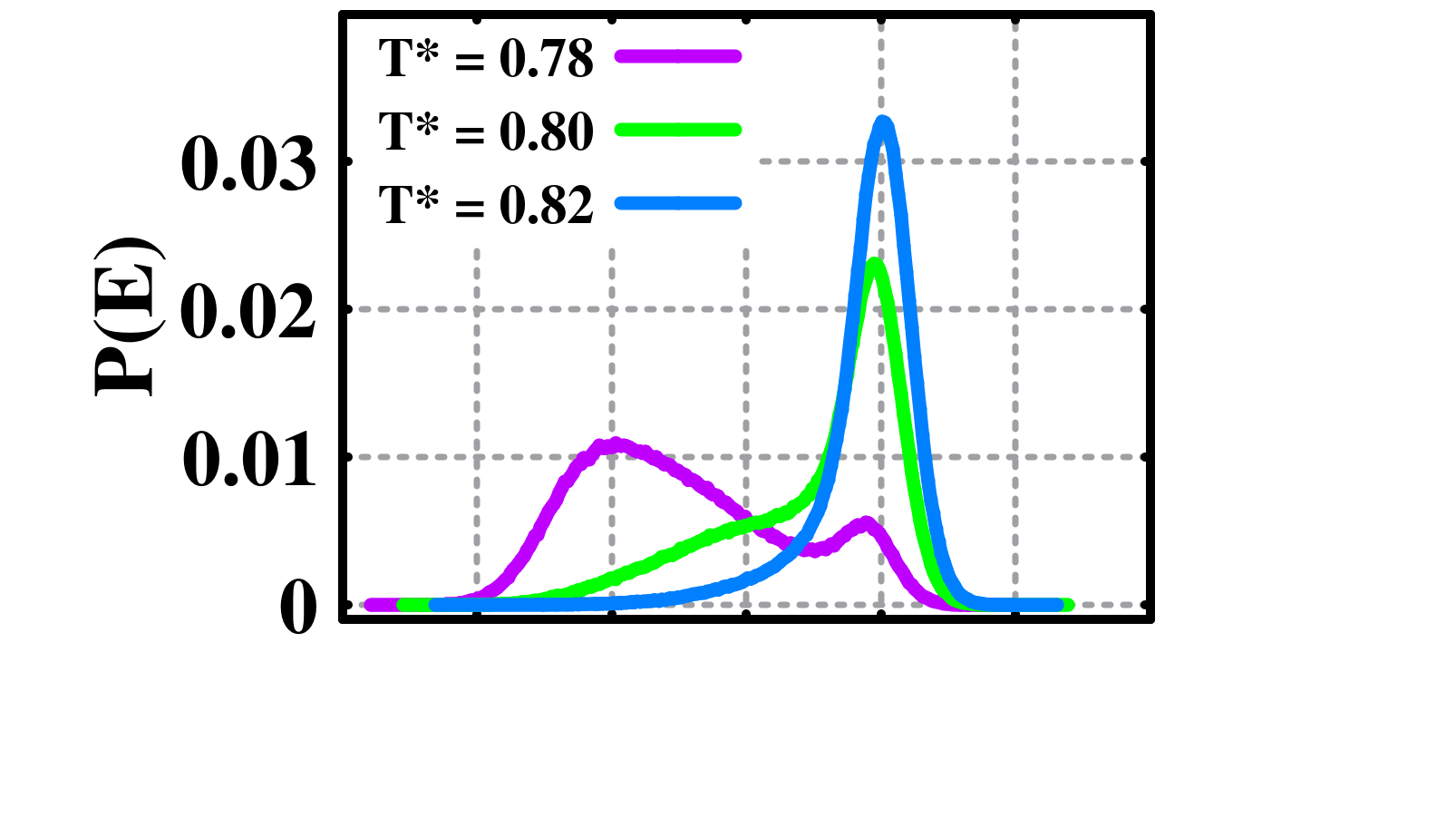}}};
\end{tikzpicture}
\begin{tikzpicture}
\node[anchor=south west,inner sep=0] (image) at (0,0){\scalebox {0.32}{\includegraphics[width=\textwidth, trim=4cm 6cm 8cm 0cm, clip=true, angle=0, page=1]{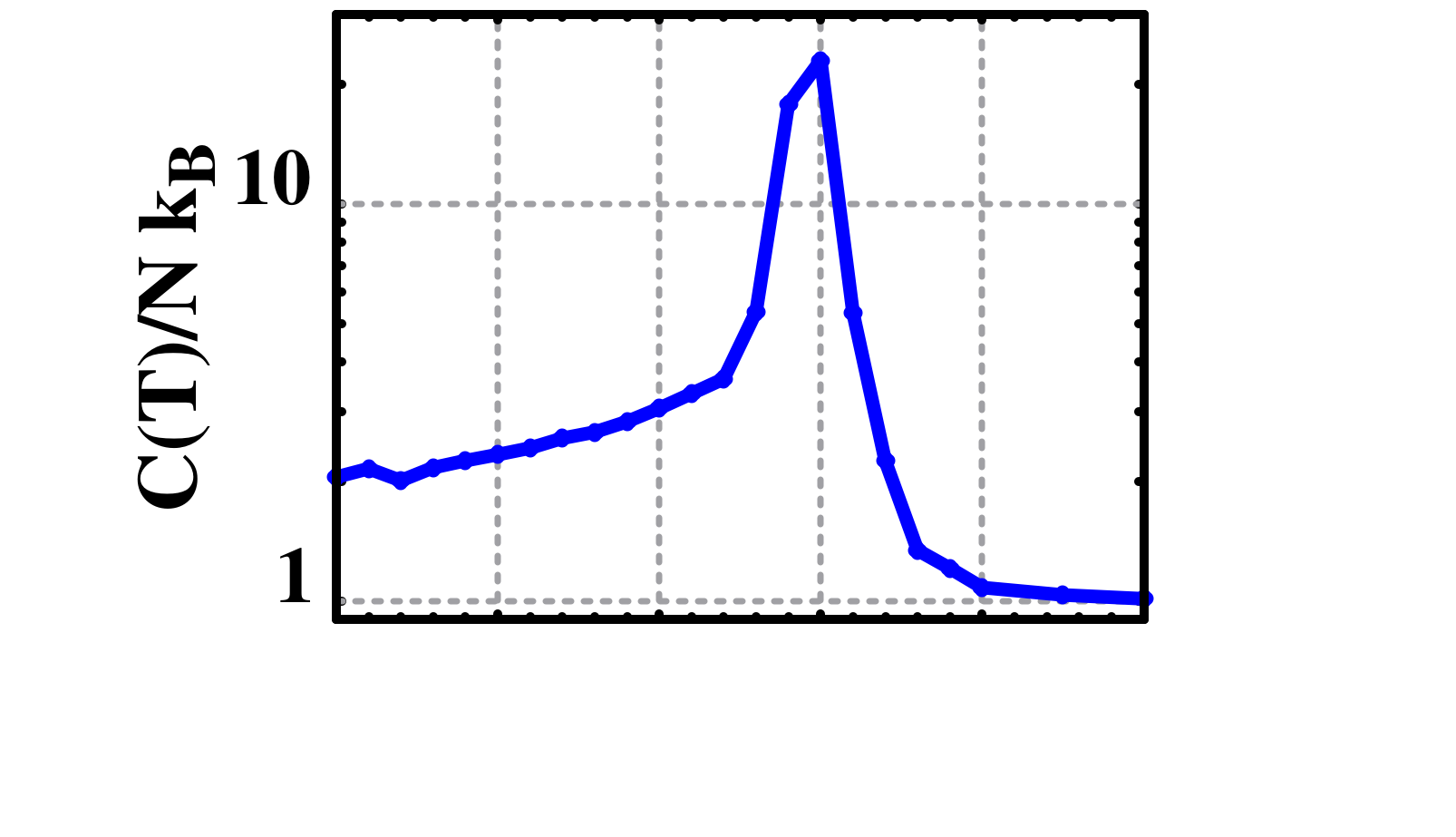}}};
\end{tikzpicture}
\begin{tikzpicture}
\node[anchor=south west,inner sep=0] (image) at (0,0){\scalebox {0.32}{\includegraphics[width=\textwidth, trim=3.5cm 6cm 8cm 0cm, clip=true, angle=0, page=1]{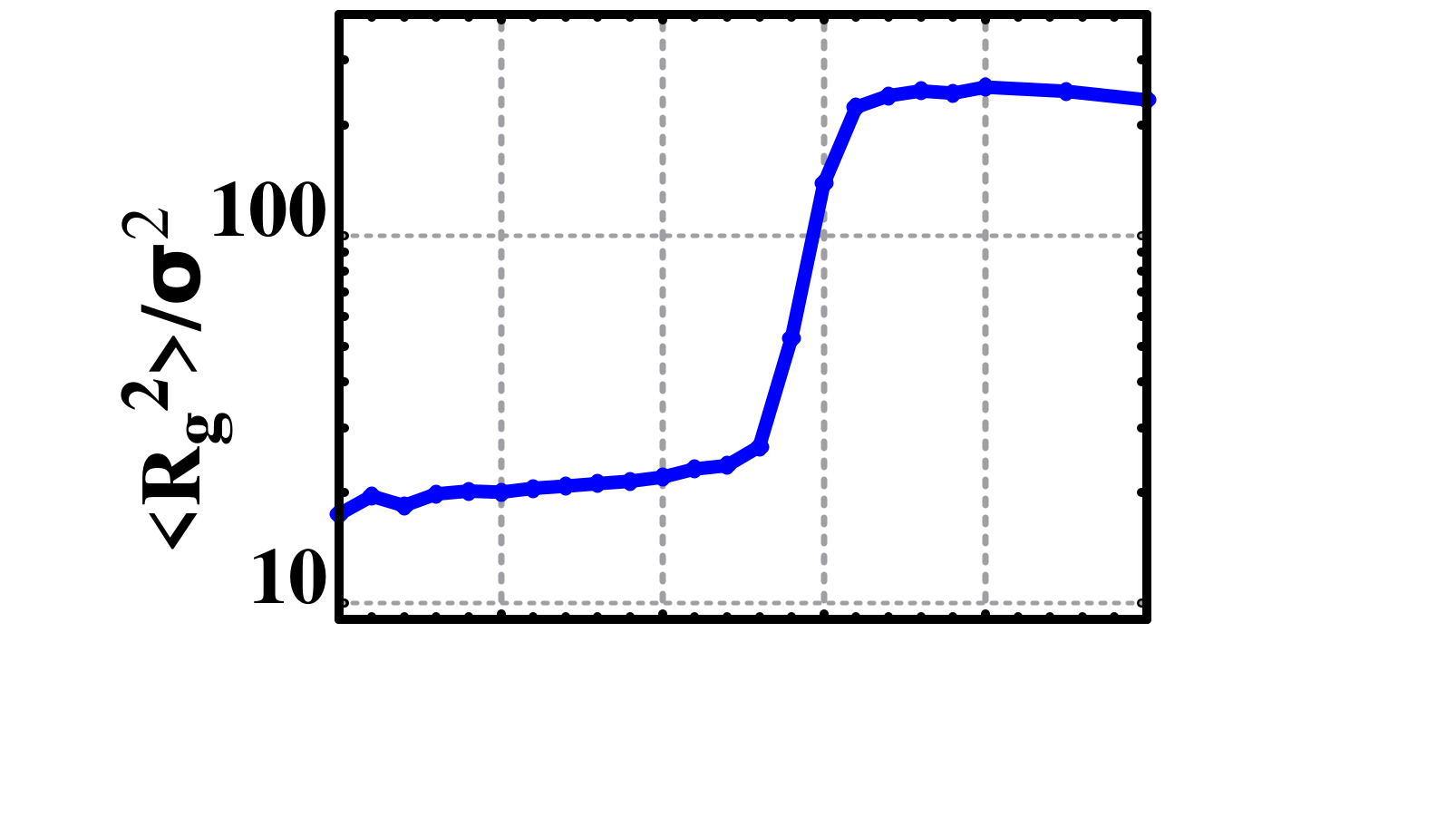}}};
\end{tikzpicture}
\begin{tikzpicture}
\node[anchor=south west,inner sep=0] (image) at (0,0){\scalebox {0.34}{\includegraphics[width=\textwidth, trim=2cm 6cm 8cm 0cm, clip=true, angle=0, page=1]{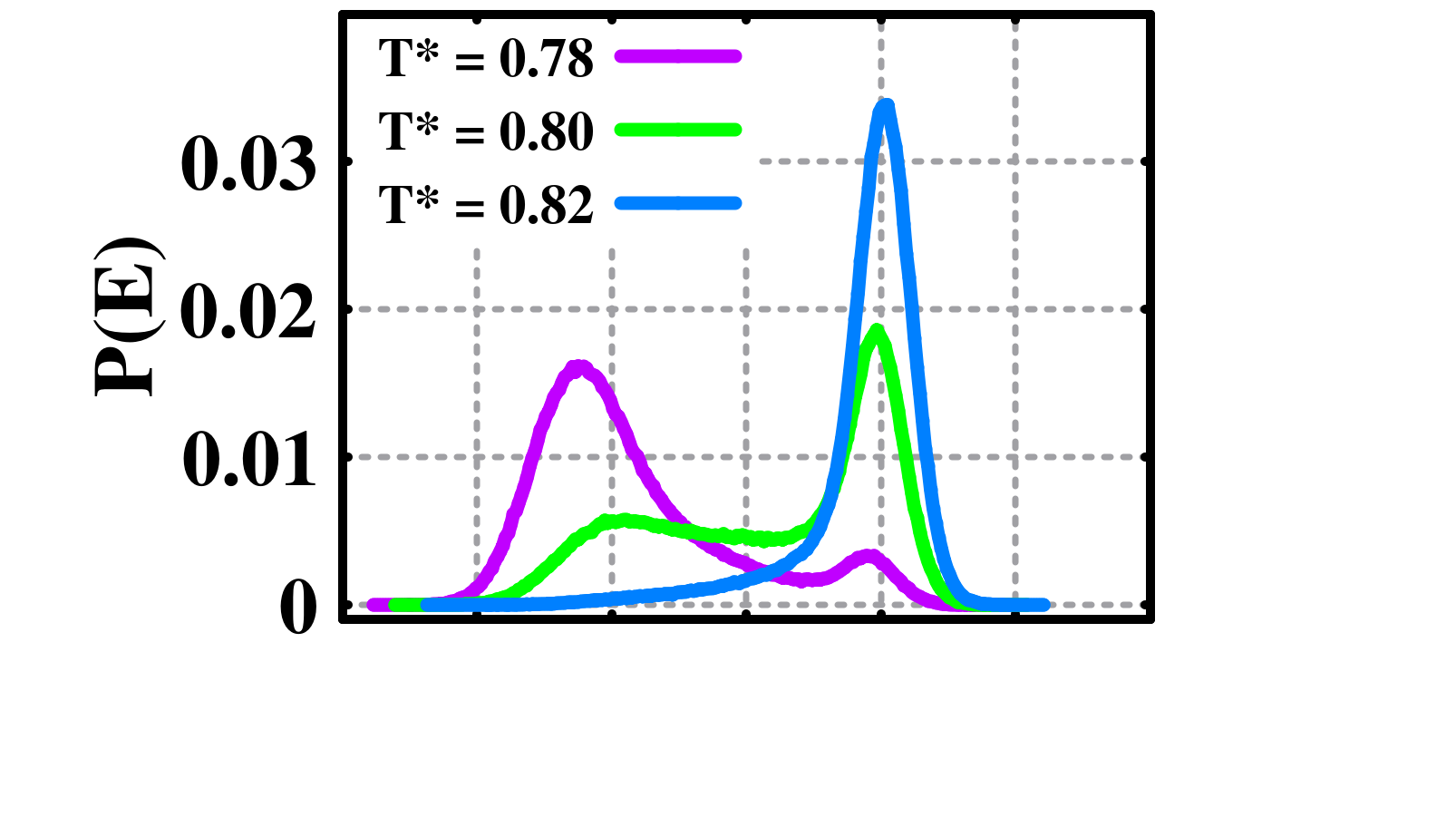}}};
\end{tikzpicture}
\begin{tikzpicture}
\node[anchor=south west,inner sep=0] (image) at (0,0){\scalebox {0.32}{\includegraphics[width=\textwidth, trim=4cm 6cm 8cm 0cm, clip=true, angle=0, page=1]{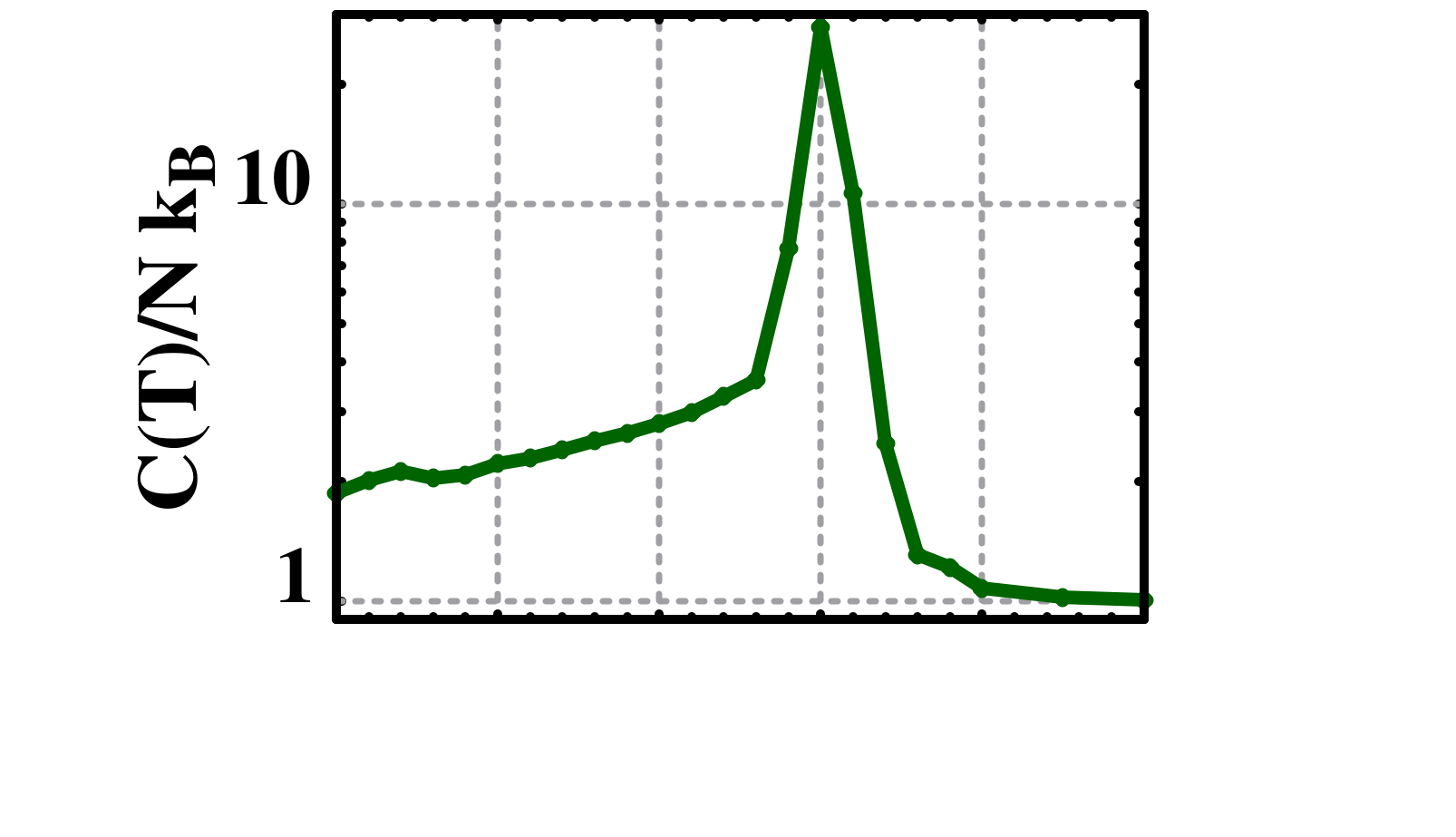}}};
\end{tikzpicture}
\begin{tikzpicture}
\node[anchor=south west,inner sep=0] (image) at (0,0){\scalebox {0.32}{\includegraphics[width=\textwidth, trim=3.5cm 6cm 8cm 0cm, clip=true, angle=0, page=1]{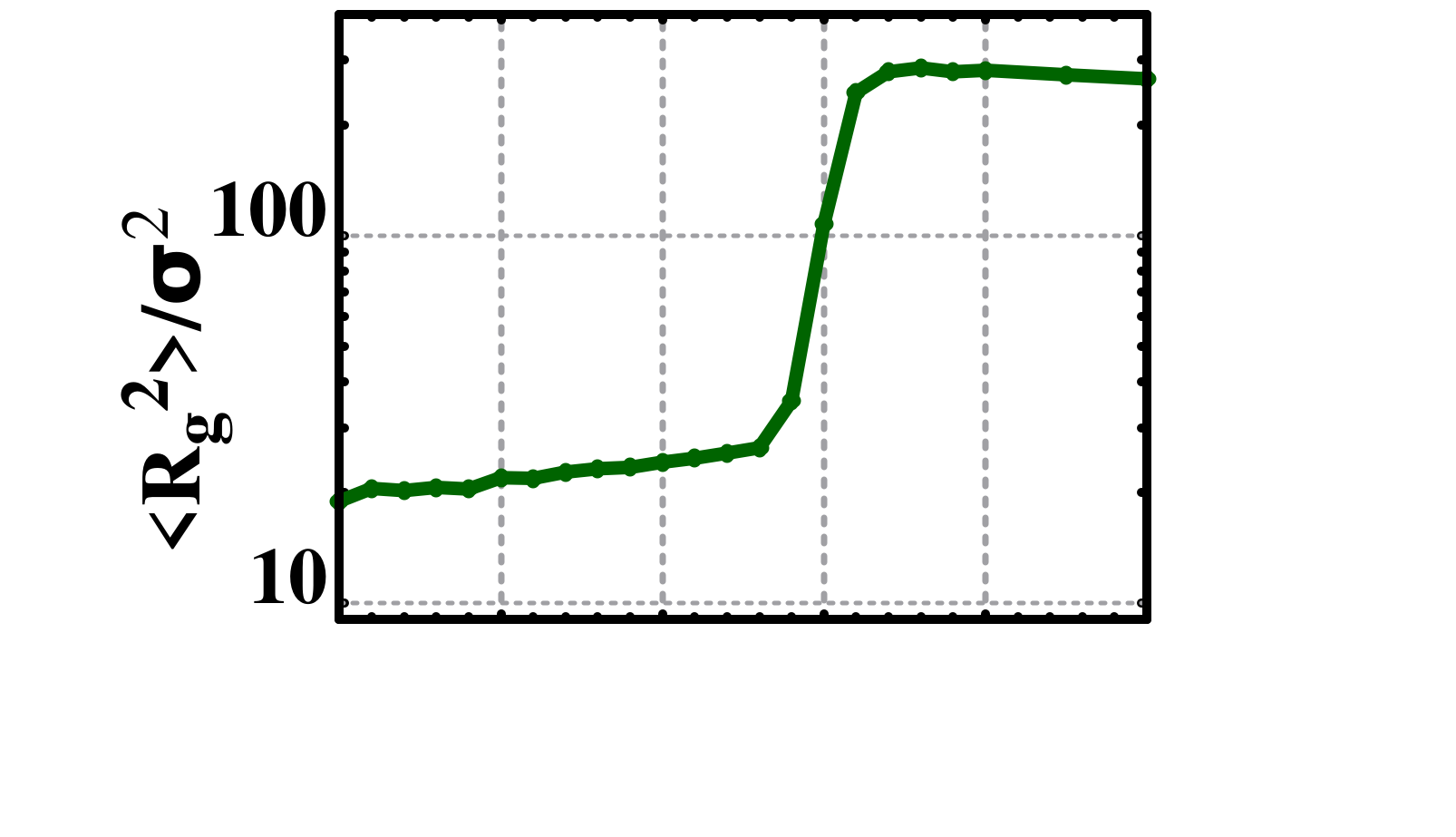}}};
\end{tikzpicture}
\begin{tikzpicture}
\node[anchor=south west,inner sep=0] (image) at (0,0){\scalebox {0.34}{\includegraphics[width=\textwidth, trim=2cm 6cm 8cm 0cm, clip=true, angle=0, page=1]{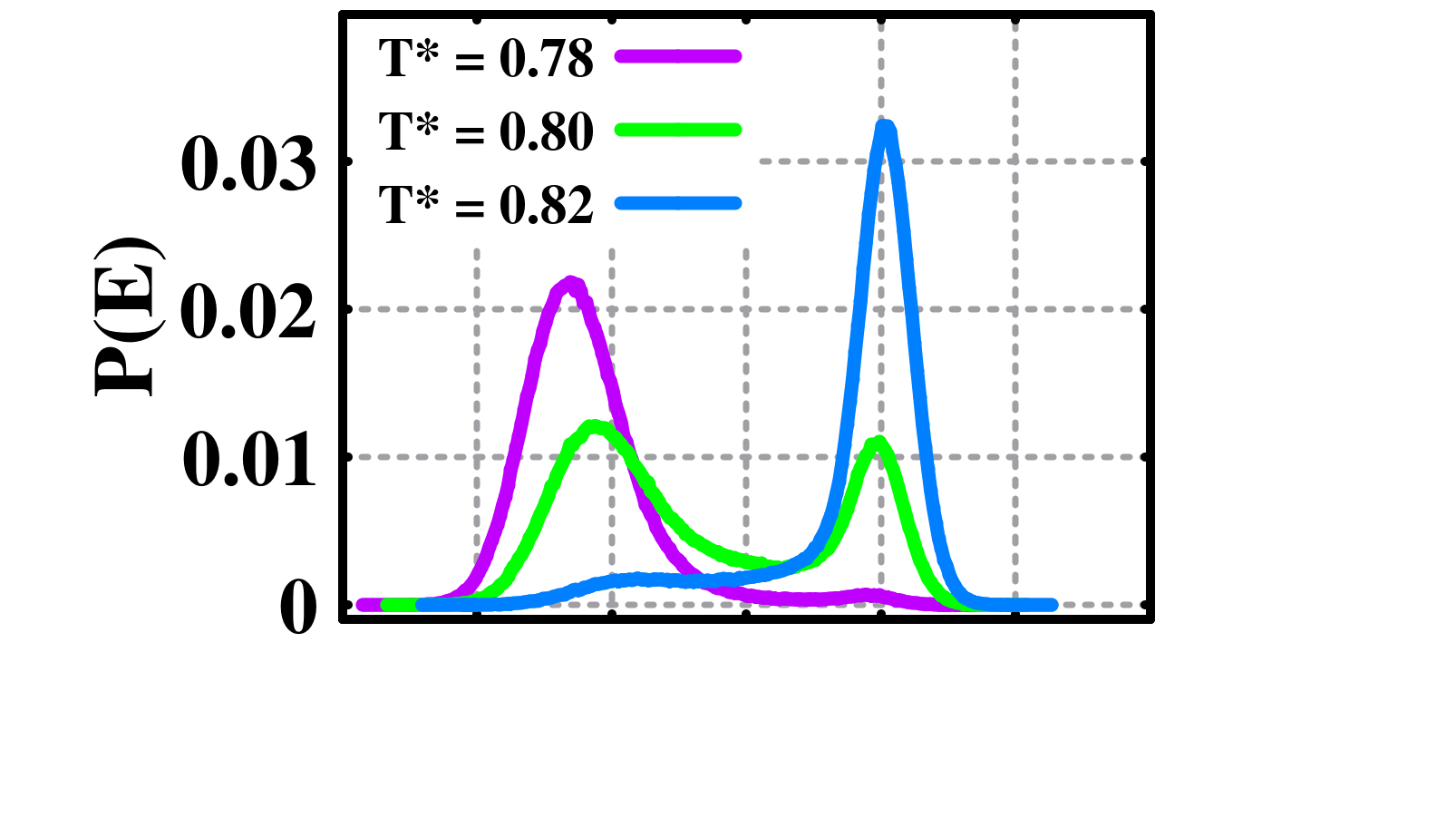}}};
\end{tikzpicture}
\begin{tikzpicture}
\node[anchor=south west,inner sep=0] (image) at (0,0){\scalebox {0.32}{\includegraphics[width=\textwidth, trim=4cm 0cm 8cm 0cm, clip=true, angle=0, page=1]{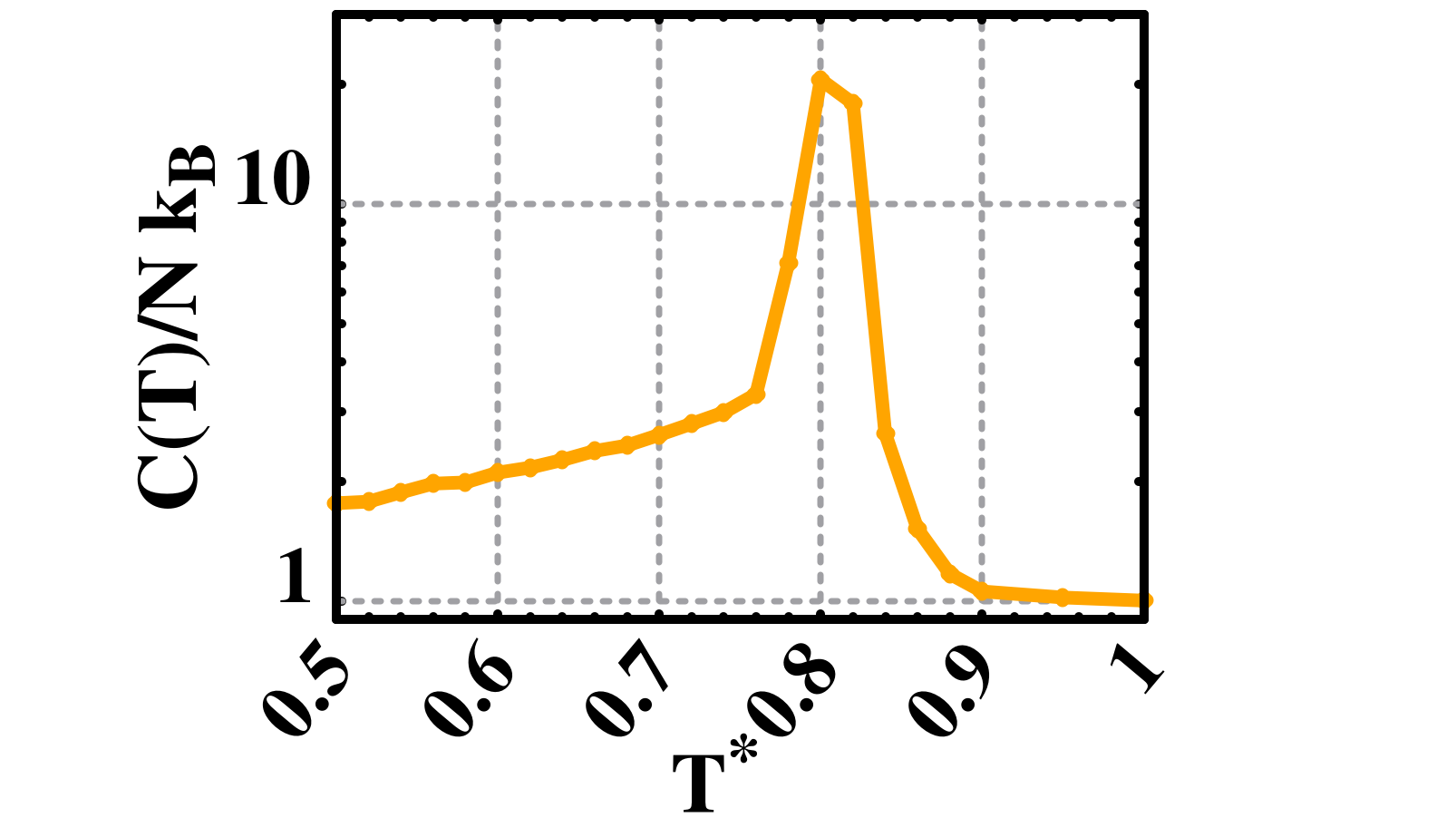}}};
\end{tikzpicture}
\begin{tikzpicture}
\node[anchor=south west,inner sep=0] (image) at (0,0){\scalebox {0.32}{\includegraphics[width=\textwidth, trim=3.5cm 0cm 8cm 0cm, clip=true, angle=0, page=1]{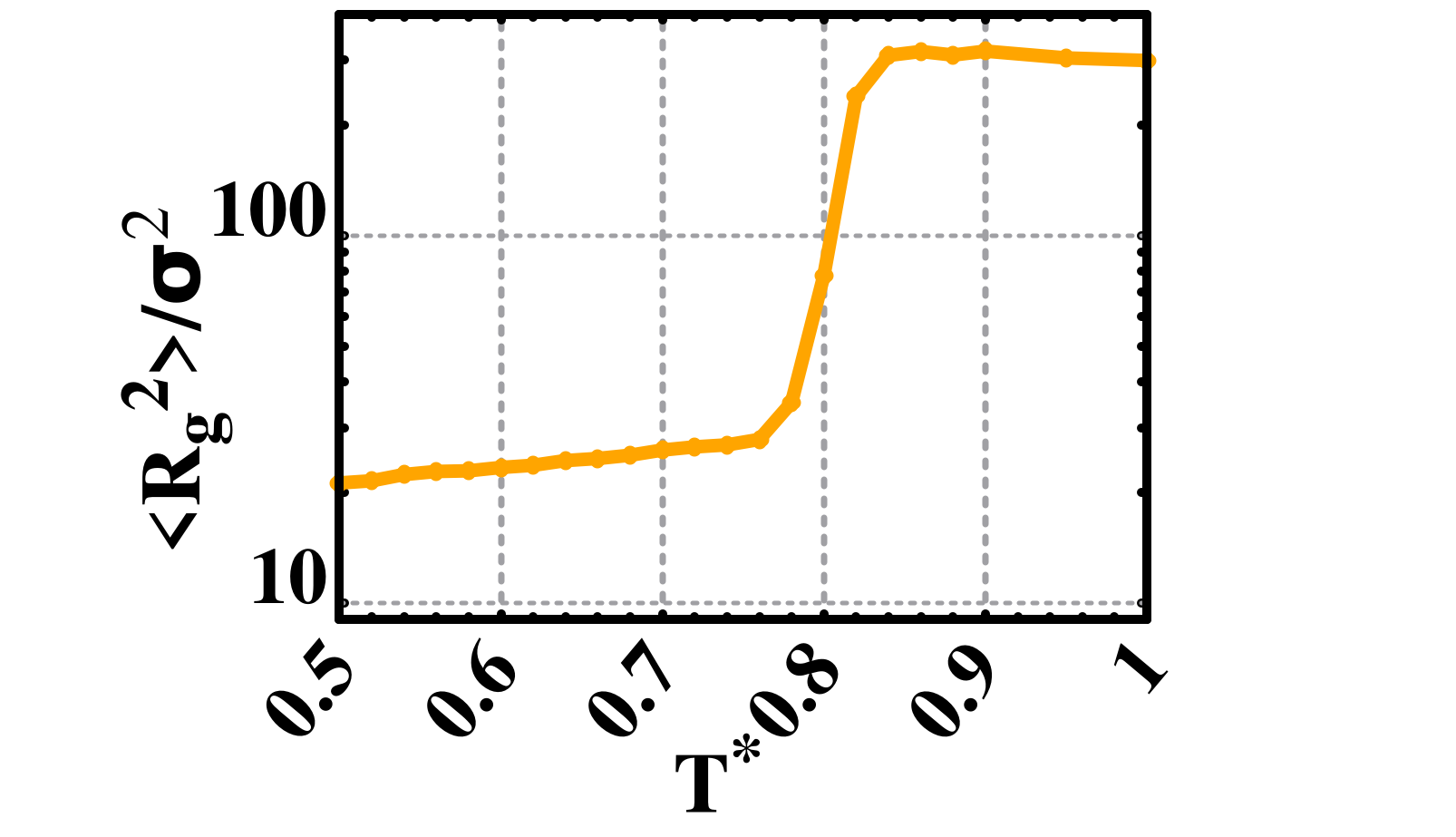}}};
\end{tikzpicture}
\begin{tikzpicture}
\node[anchor=south west,inner sep=0] (image) at (0,0){\scalebox {0.34}{\includegraphics[width=\textwidth, trim=2cm 0cm 8cm 0cm, clip=true, angle=0, page=1]{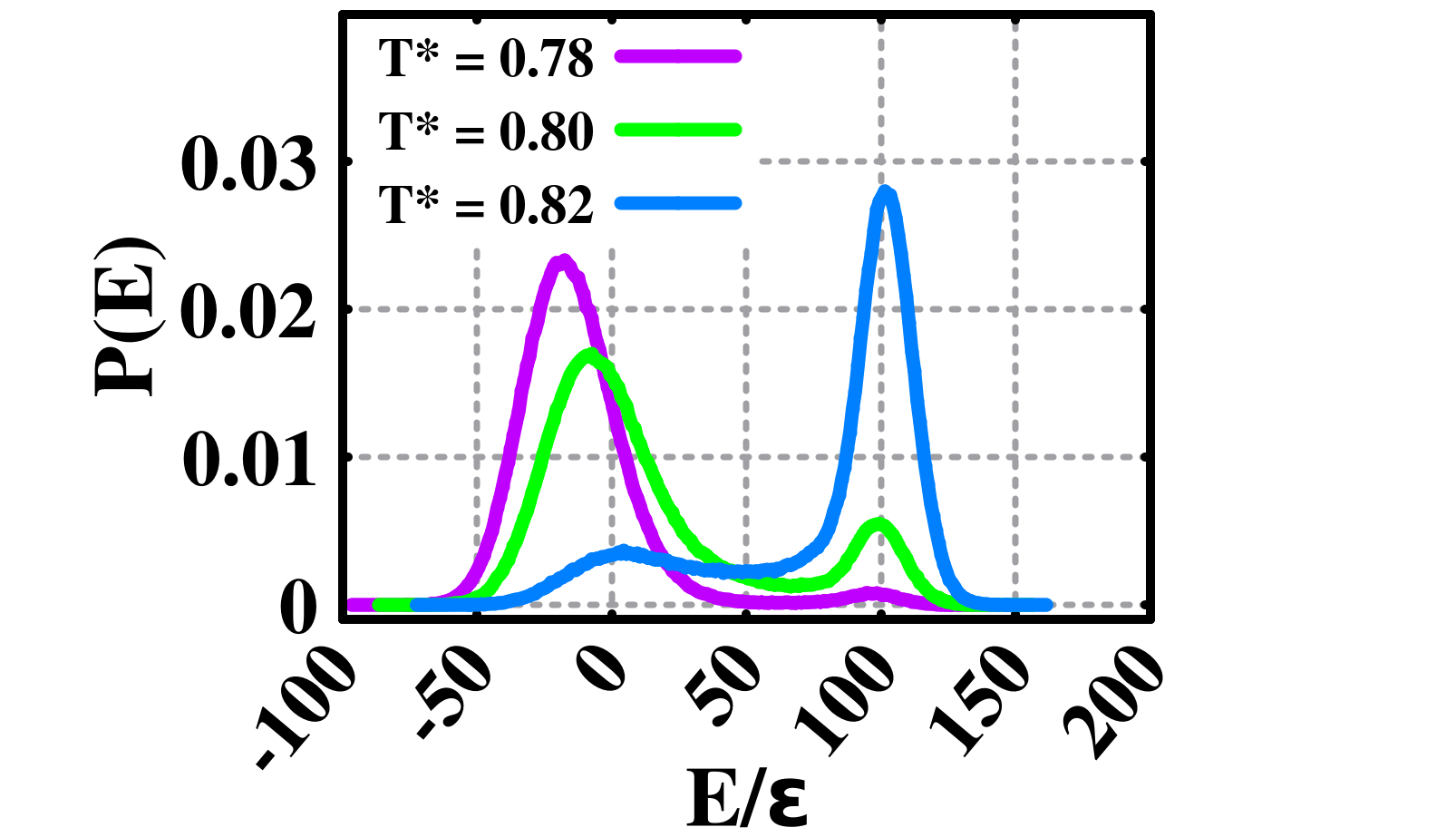}}};
\end{tikzpicture}
\caption{From top to bottom: Specific heat per monomer (first column), Radius of gyration (second column) as a function of reduced temperature $T^{*}$, and energy distribution probability (third column) evaluated at temperatures $T^{*} = 0.78$ , $T^{*} = 0.80$ and $T^{*} = 0.82$. Bending stiffness are $K = 7$ (first row) , $K = 8$ (second row) , $K = 9$ (third row) and $K = 10$ (fourth row).}
\label{fig:figs4}
\end{figure*}

The progressive shift of the direct coil-crystal transition as quantified by the specific heat per monomer in Figure \ref{fig:figs4} as stiffness increases. The mean square radius of gyration signals the abrupt collapse, and the energy probability distribution indicated the progressive onset of the double peak characteristic of the first order phase transition.
\clearpage
\subsection{Eigenvalues of the moment of inertia tensor for single chain polymer}
\label{subsec:eigenvalues}
\begin{figure*}[h]
\centering
\begin{subfigure}[b]{0.4\textwidth}
    \centering
    \includegraphics[width=7cm]{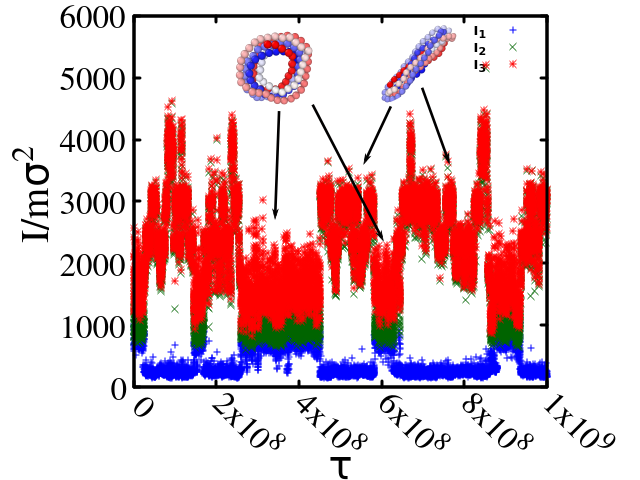}
     \caption{\label{fig:eigenvalues_a}}
    \end{subfigure}
    \qquad \qquad \qquad
    \begin{subfigure}[b]{0.4\textwidth}
    \centering
    \includegraphics[width=7cm]{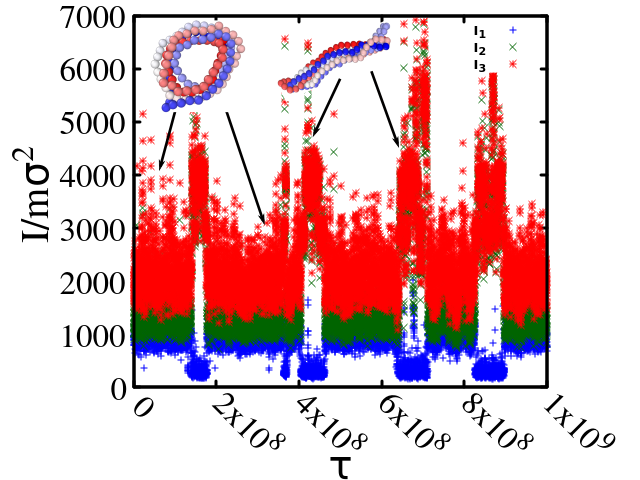}
     \caption{\label{fig:eigenvalues_b}}
    \end{subfigure}
    \caption{Eigenvalues of the moment of inertia tensor $I_1<I_2=I_3$  during the time trajectory. (a)  $K=5$, and $T^*=0.62$; (b) $K=6$, and $T^*=0.70$. Also shown here are representative snapshots of rods and toroids.}
\label{fig:figs5}
\end{figure*}

The periodic oscillations in the numerical value of the eigenvalues of the moment of inertia tensor during the trajectory (Figure \ref{fig:figs5}) is a signal of the isoenergetic switching from rods to toroids (and vice versa) in the particular region of the phase diagram indicated as S (switching) in Figure 5 of the main text. In particular, the smallest eigenvalue$I_1$ (blue points) is small for rods and large for toroids, as explained in Section \ref{sec:theory}.

\begin{figure*}[h]
    \begin{subfigure}[b]{0.3\textwidth}
    \centering
    \includegraphics[width=7cm]{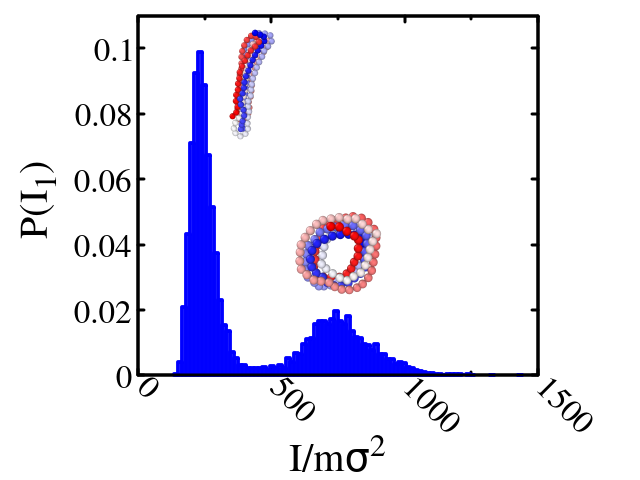}
     \caption{\label{fig:distribution_a}}
    \end{subfigure}
    \qquad \qquad \qquad
    \begin{subfigure}[b]{0.3\textwidth}
    \centering
    \includegraphics[width=7cm]{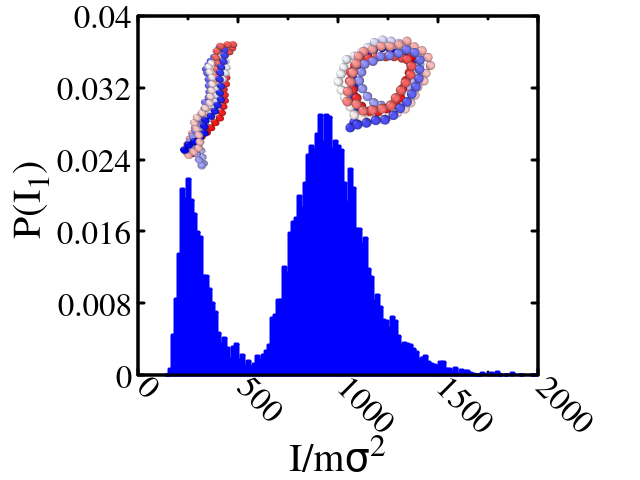}
     \caption{\label{fig:distribution_b}}
    \end{subfigure}
    \begin{subfigure}[b]{0.3\textwidth}
    \centering
    \includegraphics[width=7cm]{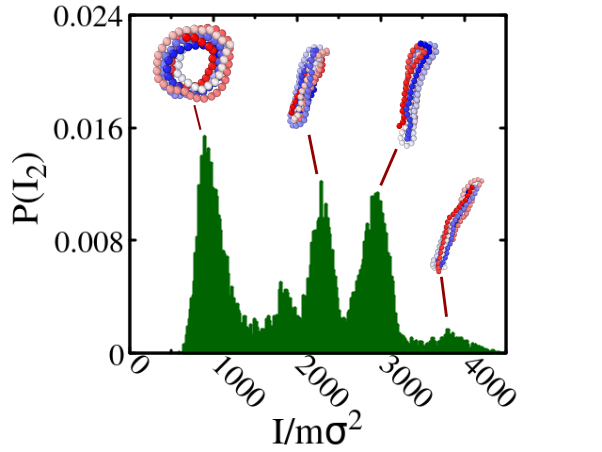}
     \caption{\label{fig:distribution_c}}    
    \end{subfigure}
    \qquad \qquad \qquad
    \begin{subfigure}[b]{0.3\textwidth}
    \centering
    \includegraphics[width=7cm]{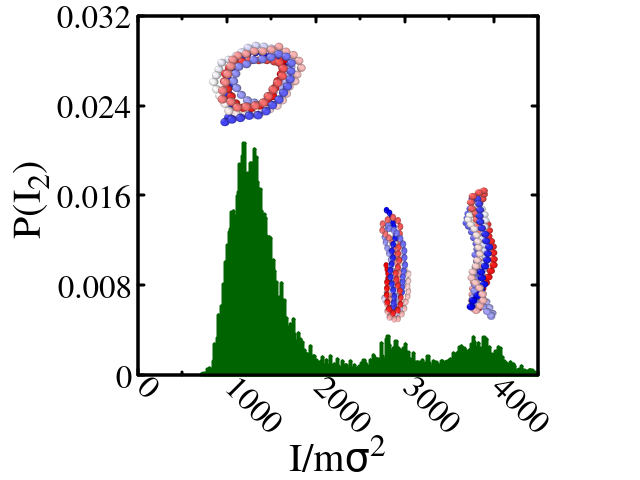}
     \caption{\label{fig:distribution_d}}    
    \end{subfigure}    
\caption{Probability distribution of $I_1$  (Panels (a) and (b)) and $I_2$ (Panels (c) and (d)) for $K=5, T^*=0.62$ (Panels (a) (c) ) and for $K=6, T^*=0.70$ (Panels (b) and (d)). Representative snapshots identify different shapes.}
\label{fig:figs6}
\end{figure*}

Probability distribution of $I_1$ and $I_2$ (Figure \ref{fig:figs6}) in the same state points as in Figure \ref{fig:figs5} that clearly illustrate that $I_1$ (and, to a lesser extent, $I_2$), are able to discriminate between rod and toroidal shapes.

\begin{figure*}[h!]
\centering
    \begin{subfigure}[b]{0.3\textwidth}
    \centering
    \includegraphics[width=6cm]{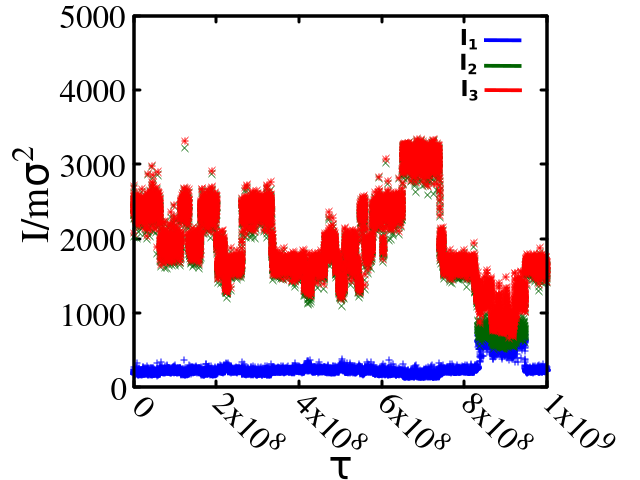}
     \caption{}\label{fig:figs7a}
    \end{subfigure}
    \qquad \qquad \qquad
    \begin{subfigure}[b]{0.3\textwidth}
    \centering
    \includegraphics[width=6cm]{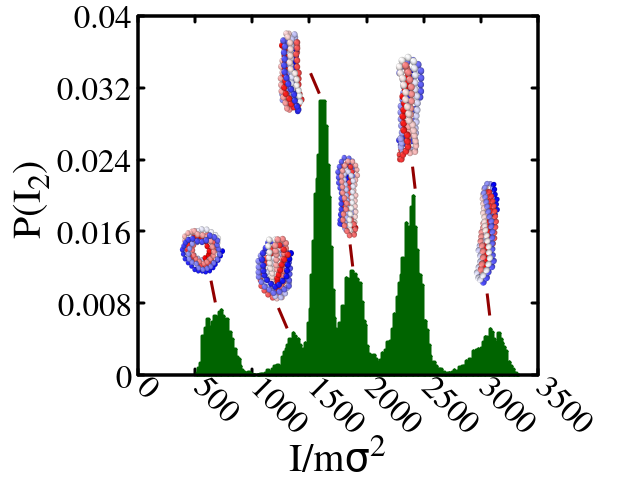}
     \caption{}\label{fig:figs7b}
    \end{subfigure}
    \qquad \qquad \qquad
    \begin{subfigure}[b]{0.3\textwidth}
    \centering
    \includegraphics[width=6cm]{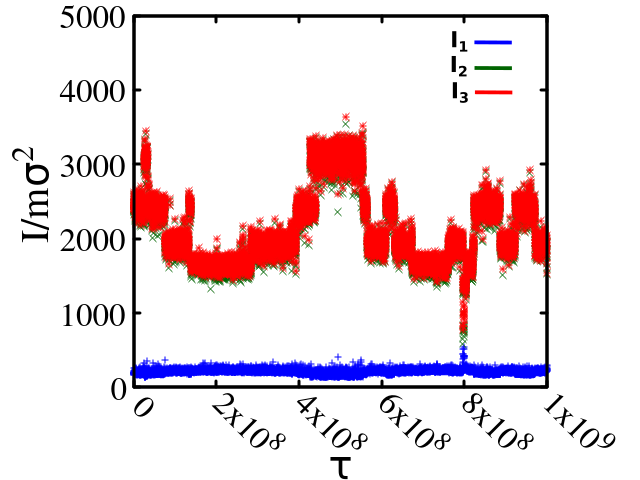}
     \caption{}\label{fig:figs7c}
    \end{subfigure}
    \qquad \qquad \qquad
    \begin{subfigure}[b]{0.3\textwidth}
    \centering
    \includegraphics[width=6cm]{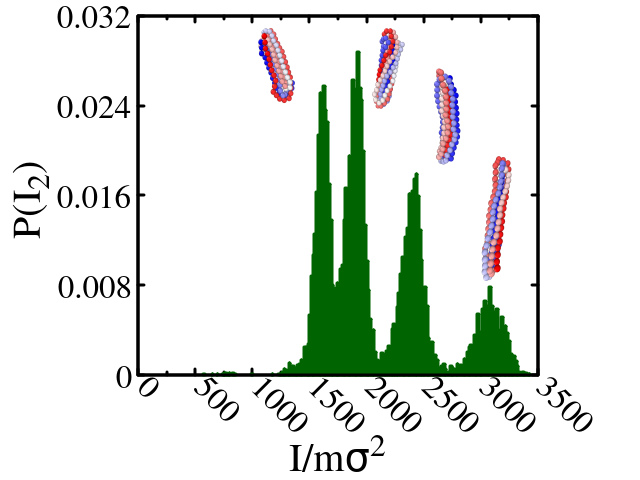}
     \caption{}\label{fig:figs7d}
    \end{subfigure}
    \qquad \qquad \qquad
    \begin{subfigure}[b]{0.3\textwidth}
    \centering
    \includegraphics[width=6cm]{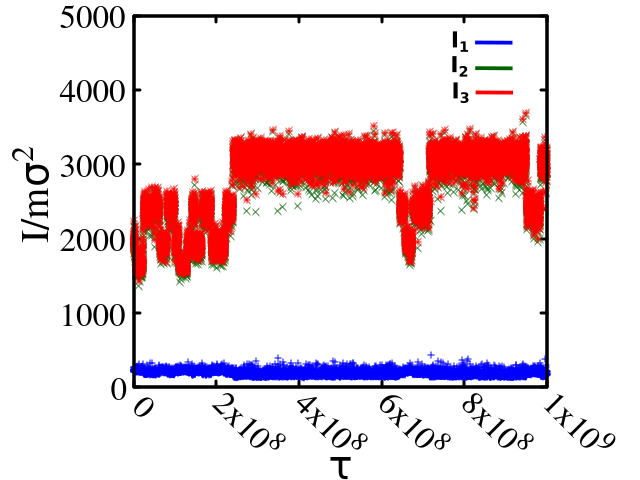}
     \caption{}\label{fig:figs7e}
    \end{subfigure}
    \qquad \qquad \qquad
    \begin{subfigure}[b]{0.3\textwidth}
    \centering
    \includegraphics[width=6cm]{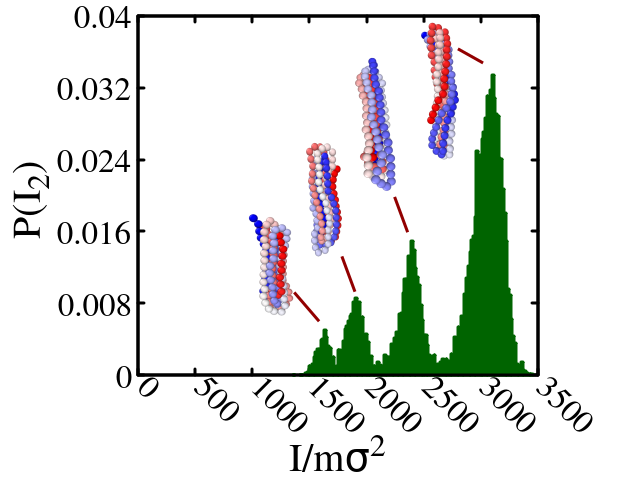}
     \caption{}\label{fig:figs7f}
    \end{subfigure}
\caption{Panels (a),(c),(e): Eigenvalues of the moment of inertia tensor $I_1<I_2=I_3$  during of the time trajectory; Panels (b),(d),(f): Probability distribution function of $I_2$. Parameter values are : (a) and (b) $K=4.2, T^*=0.52$; (c) and (d) $K=4.4, T^*=0.52$; (e) and (f) $K=4.6, T^*=0.54$.}
\label{fig:figs7}
\end{figure*}

In some region of the phase diagram of Figure 5 in the main text, the coexistence of rods of different lengths is observed. This is illustrated in Figure  \ref{fig:figs7} where the smallest eigenvalues $I_1$ stays constant at a small value (thus indicating a rod shape) but the second (degenerate) eigenvalue $I_2=I_3$ changes  signalling a coexistence between rods of different length. The multi modal profile of $P(I_2)$ suggests the presence of a discrete distribution of rods lengths. On comparing increasing values of $K$ at equal temperatures, it is possible to appreciate the shift towards larger rod lengths consistent with an increasing persistence length.
\clearpage
\subsection{Free energy landscape with $S$ as reaction coordinate for multi chains self-assembly}
\label{subsec:free}
As illustrated in the main text, the different transition lines in the phase diagram of the multi chain self-assembly (Figure 11 (b) of the main text) can also be identified through the free energy landscape using $E_{inter}$ and $S$ as reaction coordinates. In particular, the nematic order parameter is optimal to pin down the onset of the bundle phase, where different chains tend to align parallel to each other in a nematic bundle. This is reported in Figure \ref{fig:free}.  Close to the coil-globule transition the nematic order parameter $S$ remains small (Panel (a)) and two deep minima  corresponding to small and large values of $\vert E_{inter} \vert$ are present in the free energy landscape (Panel (b)) clearly indicating the coil-globule transition. Close to the globule-bundle transition, $S$ shows a marked increase to unit (Panel (c)) and the free energy landscape shows two deep minima at low $S$ and intermediate $\vert E_{inter} \vert$ (globule) and at high $S$ and high $\vert E_{inter} \vert$ (bundle). Close to the coil-bundle transition, $S$ also displays a sharp upswing (Panel (e)) and the free energy landscape presents two deep minima again at low $S$ and low $\vert E_{inter} \vert$ (coil) and at high $S$ and high $\vert E_{inter} \vert$ (bundle).
\begin{figure*}[htbp]
\centering
\begin{subfigure}[b]{0.4\textwidth}
    \centering
    \includegraphics[width=7.5cm]{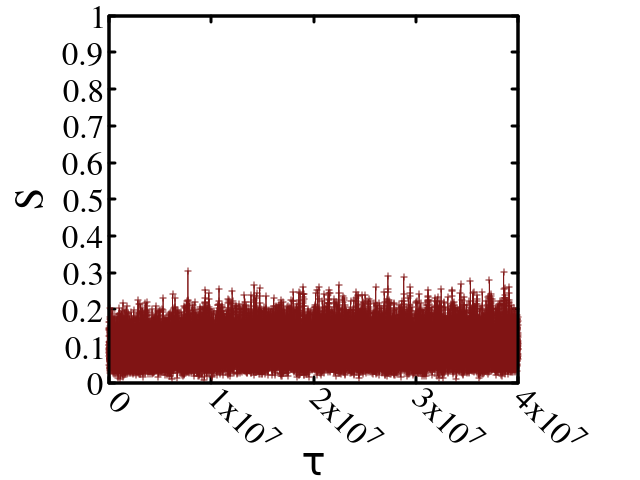}
     \caption{\label{fig:figs8a}}
    \end{subfigure}
    \qquad \qquad \qquad
    \begin{subfigure}[b]{0.4\textwidth}
    \centering
    \includegraphics[width=7.5cm]{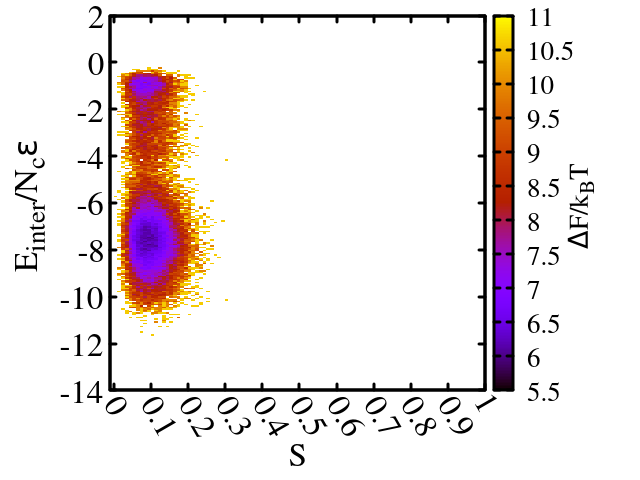}
     \caption{\label{fig:figs8b}}
    \end{subfigure}
    \qquad \qquad \qquad
    \begin{subfigure}[b]{0.4\textwidth}
    \centering
    \includegraphics[width=7.5cm]{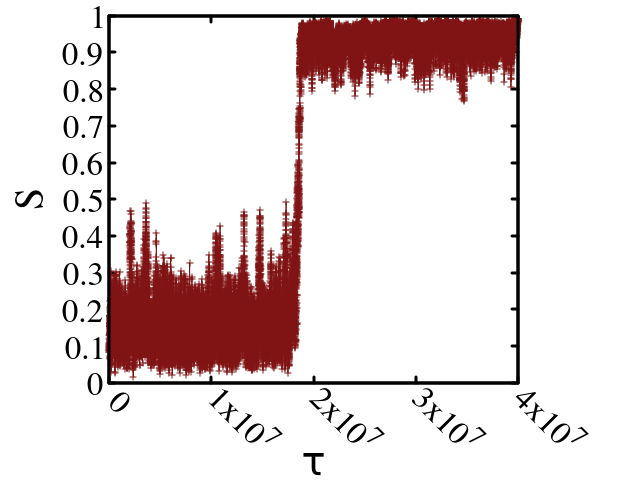}
     \caption{\label{fig:figs8c}}
    \end{subfigure}
    \qquad \qquad \qquad
    \begin{subfigure}[b]{0.4\textwidth}
    \centering
    \includegraphics[width=7.5cm]{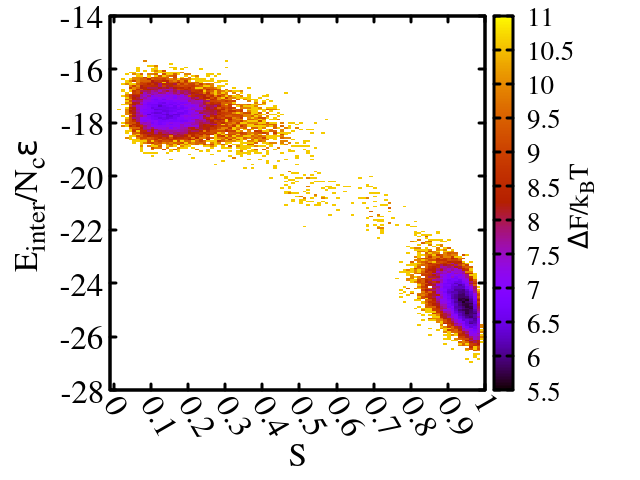}
    \caption{\label{fig:figs8d}}
    \end{subfigure}
    \qquad \qquad \qquad
    \begin{subfigure}[b]{0.4\textwidth}
    \centering
    \includegraphics[width=7.5cm]{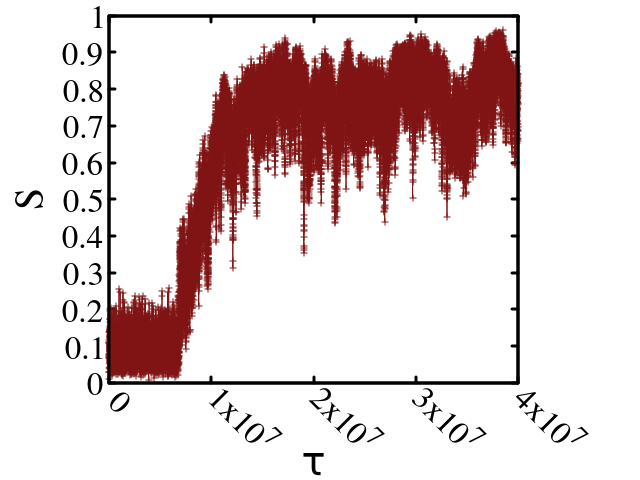}
     \caption{\label{fig:figs8e}}
    \end{subfigure}
    \qquad \qquad \qquad
    \begin{subfigure}[b]{0.4\textwidth}
    \centering
    \includegraphics[width=7.5cm]{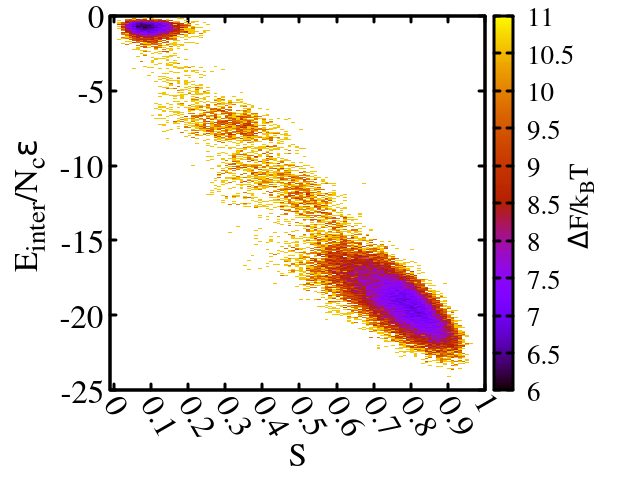}
     \caption{\label{fig:figs8f}}
    \end{subfigure}
\caption{Panels (a),(c),(e): Nematic order parameter $S$ as a function of time; Panels (b),(d),(f): Free energy landscape using inter chain energy and nematic order parameter $S$ close to the following transition lines: (a) and (b) Coil-globule transition at $K = 4$, $T^* = 0.80$, (c) and (d) Globule - bundle transitions at  $K = 4$, $T^* = 0.68$, (e) and (f) Coil-bundle transition at  $K = 8$, $T^* = 0.82$.  }
\label{fig:figs8}
\end{figure*}
\clearpage
\bibliography{macromolecules2024}

\end{document}